\documentclass{aa}  

\usepackage{graphicx}
\usepackage{xspace}
\usepackage{txfonts}
\usepackage{upgreek}
\usepackage{siunitx}

\usepackage{natbib}
\usepackage{gensymb}
\usepackage[
    colorlinks=true,
    citecolor=blue,
    linkcolor=black,
    urlcolor=blue
]{hyperref}
\newcommand{\xmm}{\emph{XMM-Newton}\xspace}
\newcommand{\micron}{\(\upmu\)m\xspace}
\def\angstrom{\leavevmode\hbox{Å}}

\begin{document}

   \title{The optical constants and grain sizes of interstellar dust measured directly using the dust scattered x-ray halo of GRB\,221009A}
    
    \titlerunning{Grain properties from the x-ray scattering halo of GRB\,221009A}

    \author{Albert Sneppen\inst{\ref{addr:DAWN},\ref{addr:jagtvej}} \and
    Darach Watson\inst{\ref{addr:DAWN},\ref{addr:jagtvej}}
    }

    \institute{Cosmic Dawn Center (DAWN)\label{addr:DAWN}
    \and
    Niels Bohr Institute, University of Copenhagen, Blegdamsvej 17, K{\o}benhavn 2100, Denmark\label{addr:jagtvej} 
    }
   
   \date{Received \today; accepted }

 
    \abstract
    {X-ray scattering is a powerful probe of the grain size distribution of interstellar dust. Bright, transient sources are excellent tools for this, since they fade rapidly, leaving only the expanding scattered x-ray halo.}
    {We analyse the dust-scattered X-ray halo data of the unprecedentedly bright $\gamma$-ray burst, GRB\,221009A, to measure the grain size distribution of dust in the Galaxy as well as the complex refractive index, $m$, and use these results to infer the likely dust composition.}
    {GRB\,221009A produced 20 distinct rings as observed with follow-up observations of the GRB afterglow and scattering halo with \emph{XMM-Newton}'s EPIC camera. We use anomalous diffraction theory to model the ring's brightness as a function of angle.}
    {We constrain the complex refractive index, $m=n+ik$ at several x-ray energies, finding $k_{1\,\mathrm{keV}}= (2.7 \pm 0.7)\times10^{-4}$ and $1-n_{1\,\mathrm{keV}}= 0.0009 \pm 0.0002$, strongly inconsistent with the commonly employed assumptions of the Rayleigh-Gans approximation. These results lie in the expected range for interstellar dust compositions dominated by carbon, magnesium silicates, and iron. The absorption results suggest a substantial mass fraction of iron at $35\pm7\%$. The \citet[MRN]{Mathis1977} distribution fit returns a maximum grain radius, $a_{\rm max}=0.24\pm 0.01$\,\micron; all fits strongly rule out models with $\sim0.4$\,\micron grains for this sightline. The soft x-ray spectrum of the prompt GRB can also be inferred from the fitting, with the best-fit providing a spectral slope that is consistent with the slope of the low energy side of the best-fit Band model of the directly measured prompt emission. Forcing a different grain size or composition than the best fit results in an inferred prompt spectrum different to the observed prompt emission.}
    {We have directly measured the grain size distribution and refractive index of the interstellar dust. There are very few grains larger than about $\sim0.3$\,\micron in radius. The refractive index is consistent with standard average dust compositions, showing that x-ray scattering is an effective tool to measure interstellar dust optical properties.}
    
    \maketitle
%
\section{Introduction}
Bright Galactic x-ray sources produce haloes around them due to small-angle scattering off dust grains in the Milky Way's (MW) interstellar medium (ISM) \citep{Overbeck1965,Rolf1983}. The energy and angular dependence of the haloes provide a means to determine the properties of the dust grains \citep{Mauche1986,Predehl1995,Draine2003}. However, the uncertainty in the distances to Galactic sources, as well as the necessary brightness of the central sources often impedes really high-quality measurements of the angular dependence, while very high signal-to-noise ratio observations with good spectral resolution are required to constrain the energy dependence very well \citep{Costantini2022}. 

Transients allow the central source brightness problem to be circumvented since the central source fades before the halo is seen, resulting in rings that are observed to expand away from a relatively faint central source on timescales of days to weeks. Such rings have been seen around Galactic x-ray binaries \citep{Smith2002,Heinz2015,Beardmore2016}, a magnetar \citep{Tiengo2010} and gamma-ray bursts \citep[GRBs,][]{Vaughan2004,Vaughan2006,Tiengo2006,Pintore2017,Tiengo2023}. GRBs in particular also offer the benefit of being very far away with respect to the MW ISM, allowing the distances to the dust-scattering layers to be determined precisely. GRB x-ray scattering rings therefore permit surprisingly good constraints on the dust's line-of-sight spatial distribution, column densities, grain sizes, and even, as we demonstrate in this paper, optical properties, provided the GRB is sufficiently bright and observed for long enough \citep[e.g.][]{Vaughan2004,Watson2006}. 

The composition of interstellar dust grains is a focus of much research, including the debated solid form of interstellar iron \citep[e.g.][]{Dwek2016,Zhukovska2018}. High resolution x-ray spectra permit examination of the x-ray edges, enabling the study of extended x-ray absorption fine structures \citep[e.g.][]{devries2009}, compositional constraints \citep[e.g.][]{Zeegers2019,Psaradaki2023}, likely oxidation states \citep{Corrales2024}, etc. As x-rays are semitransparent to dust grains, this wavelength regime offers a unique window into the solid phase of the ISM.

GRB\,221009A is the brightest GRB ever observed and occurred at low Galactic latitude \citep{Williams2023,Malesani2023}, which produced an exceptional set of expanding x-ray rings captured by \emph{Swift}-XRT and \xmm \citep{Tiengo2023}. These data have been used to constrain the radial distance to various dust concentrations \citep[e.g.][]{Vasilopoulos2023}, dust column-densities \citep[e.g.][]{Zhao2024}, and to infer the fluence of the GRB prompt emission in the soft x-ray \citep{Tiengo2023,Vaia2025}. 

\begin{figure*}
    \centering
    \includegraphics[width=0.85\linewidth]{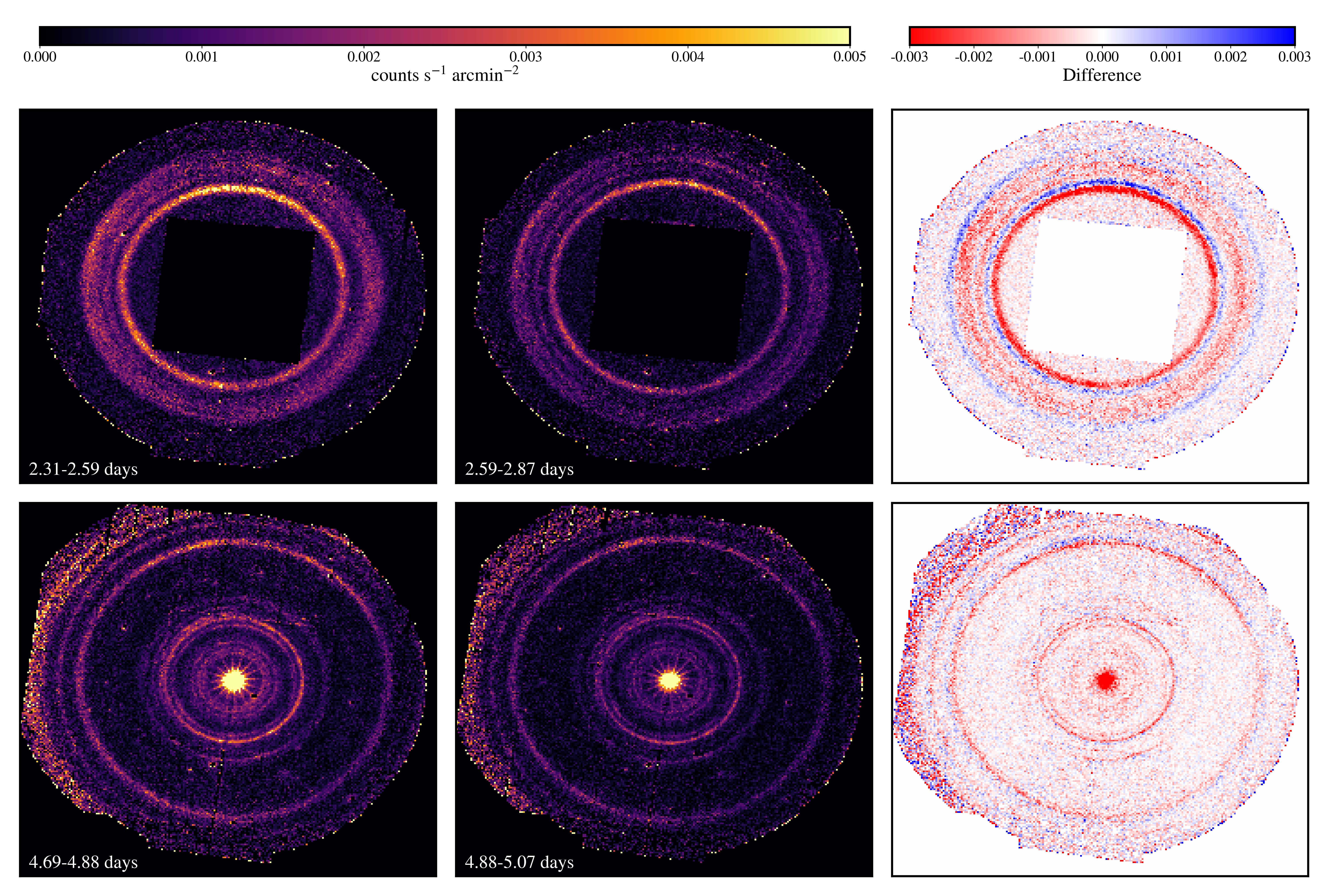}
    \caption{ \emph{Left and centre:} exposure-corrected 0.5--4\,keV images observed about six hours apart. \emph{Right:} difference images highlighting the short timescale evolution over six hours with a blue-red pattern showing rings undergoing outward motion and subtle fading. The top row shows data from the first epoch, around 2.3 days, from EPIC's MOS1 and MOS2 cameras, while the bottom row shows the second epoch data, from around 4.7 days from EPIC MOS1, MOS2, and pn. } 
    \label{fig:EPN-MOS}
\end{figure*}

\begin{figure*}
    \centering
    \includegraphics[width=0.45\linewidth,viewport=22 10 500 412 ,clip=]{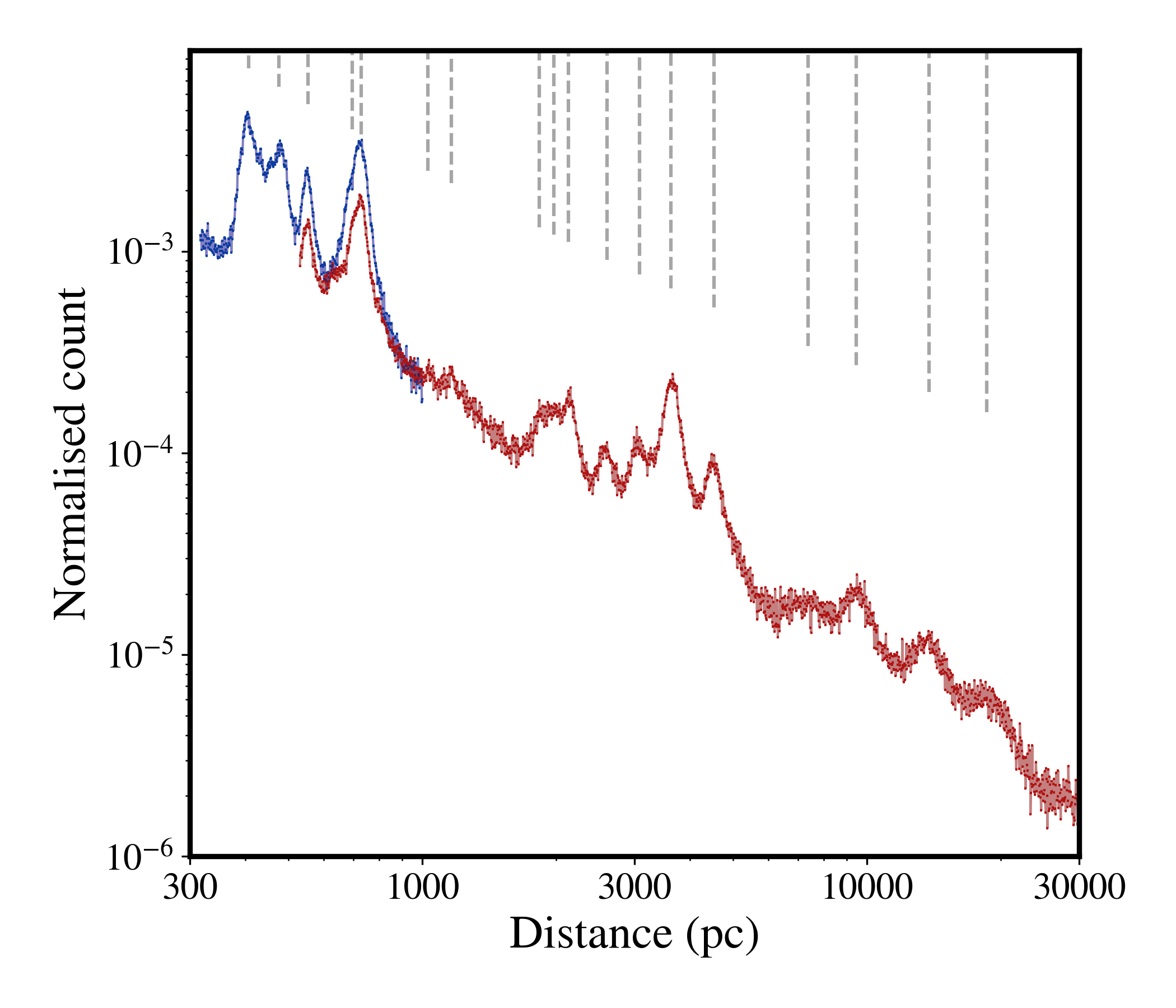}
    \includegraphics[width=0.45\linewidth,viewport=22 10 500 412 ,clip=]{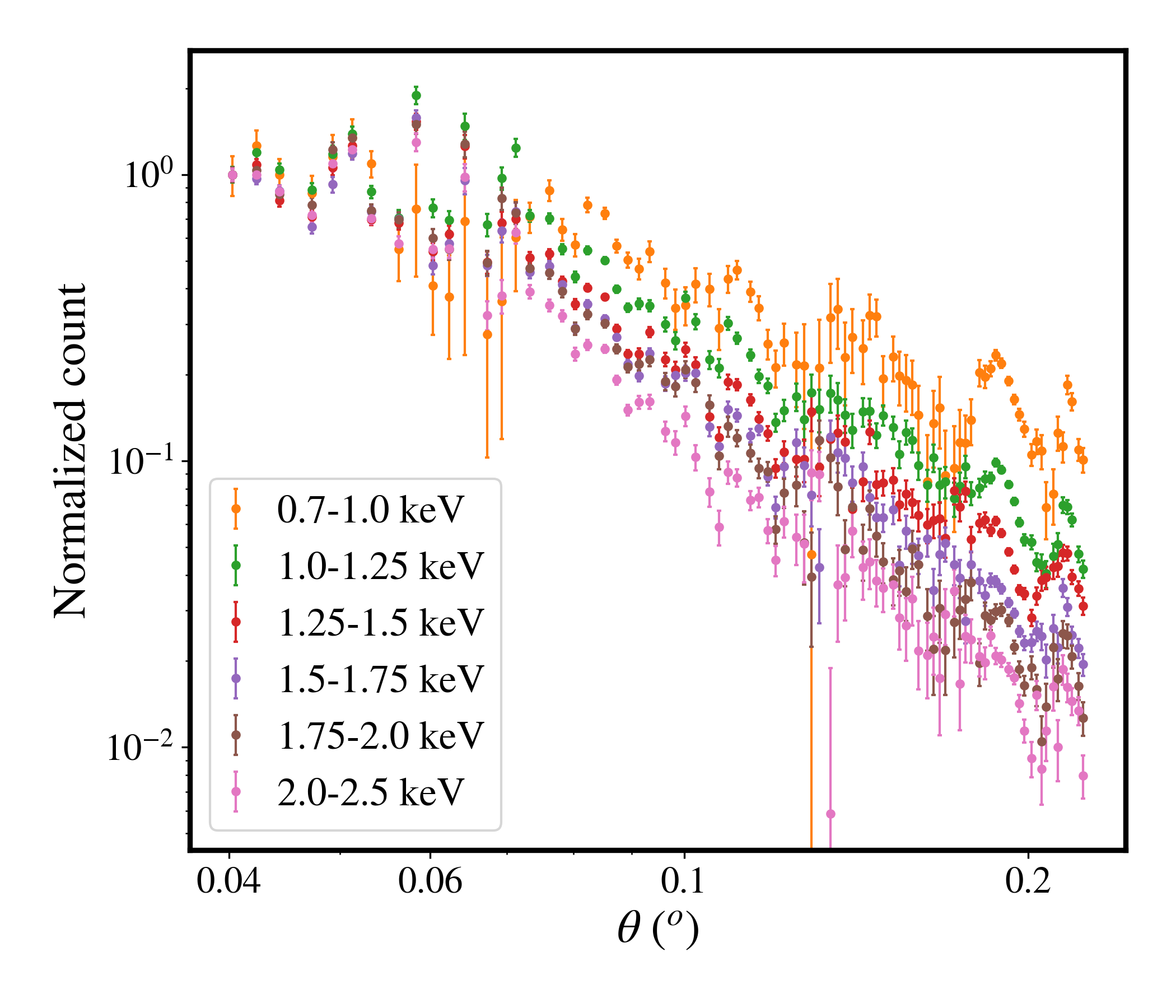}
    \caption{ X-ray counts as a function of the two major derived dimensions of our data-set -- dust distance (\emph{left}) and scattering angle (\emph{right}). The left panel shows the normalised histogram of counts in logarithmic bins of distance, i.e.\ angular distance from the GRB position transformed using $D=2ct\theta^{-2}$ (see Eq.\,\ref{eq:time_angle_distance}). Epochs~1 and 2 are shown with blue and red colours respectively, while distinct rings (i.e.\ dust-sheets associated with increased scattering) are indicated with dashed grey lines.
    The right panel shows the angular dependence of the counts, where we have marginalised over the distances i.e.\ by summing counts at various distances weighted by the inverse of the inferred radial dust density at that distance. This highlights the typical fading of rings with increasing angle and that for harder photon energies the rings fade more rapidly with angle.}
    \label{fig:EPN-MOS2}
\end{figure*}

In this analysis we take a different approach, using these observations to constrain the fading of the rings at various energies. This strongly constrains the material properties of the grains including the size distribution and, as far as we are aware, for the first time a direct determination of the complex refractive index of the MW's interstellar dust at x-ray energies. 


%
\section{Data and Methods}

\subsection{Data reduction}\label{sec:data_reduction}
\begin{figure}[h!]
    \centering
    \includegraphics[width=0.97\columnwidth,viewport=22 150 660 1630 ,clip=]{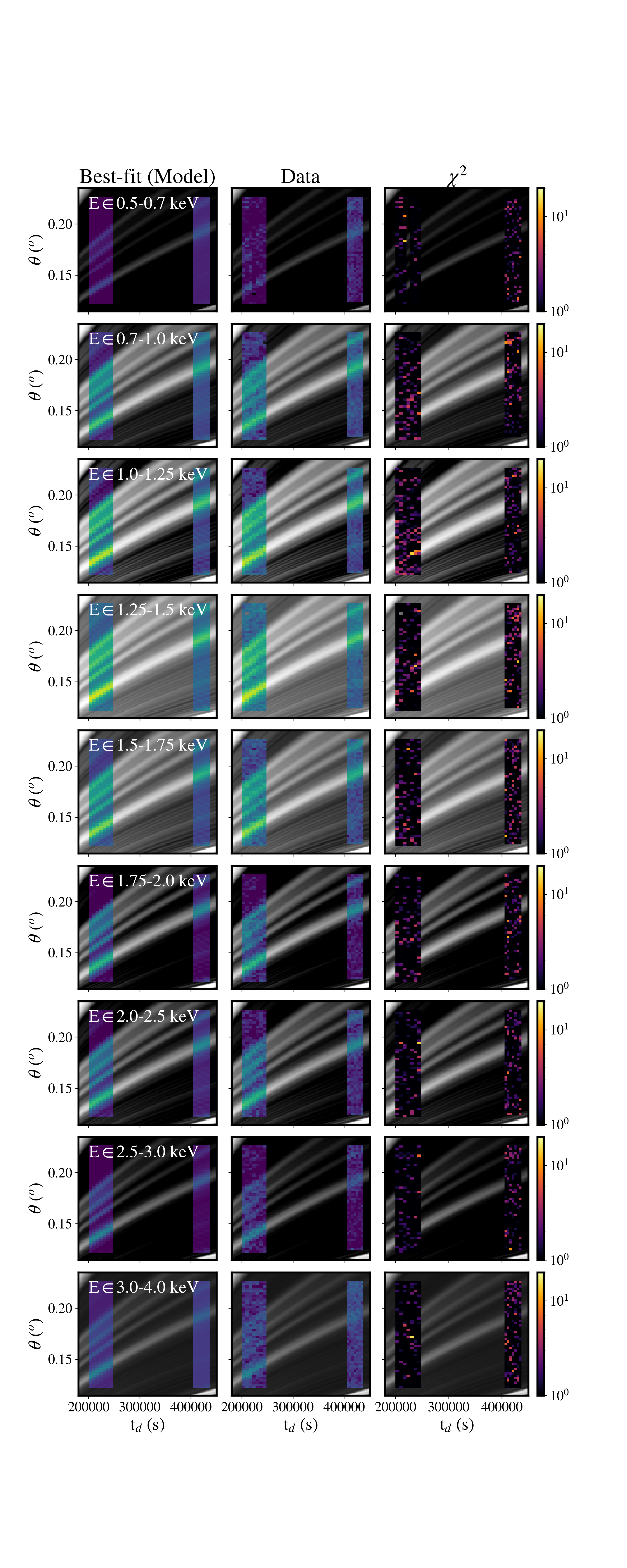} 
    \caption{X-ray halo model intensity (first column), observed intensity (middle column) and $\chi^2$-comparison (last column) as a function of angle ($\theta$) and  post-GRB time ($t_d$) in different energy-bands. The rings expand with time and, particularly at higher energies, fade with increasing angle. 
    An MRN grain-size distribution (with parameters summarised in Fig.~\ref{fig:posterior_refractive}) provide a good description of the fading of haloes within and between epochs. The pixels associated with large deviations from the model (i.e.\ large $\chi^2$) are distributed with no particular correlation to the position of the dust rings.}
    \label{fig:MRN_model}
\end{figure}

The evolving x-ray haloes were observed with the \xmm satellite at several epochs \citep[for details see][]{Tiengo2023}: Epoch~1 was on 11 October 2022, at 2.3--2.9 days post-GRB, and only the peripheral CCDs of the EPIC MOS cameras were used for imaging giving an observed angular region of $0.1\degree\lesssim\theta\lesssim0.2\degree$, which probed scattering on dust layers at distances of 300--1000\,pc at these times; Epoch~2 was on 14 October 2022 at 4.7--5.1 days post-GRB, where both the EPIC pn and MOS cameras were used for imaging with an angular region of $\theta<0.2\degree$, which at this time probed scattering on dust 500--20\,000\,pc. The prominent dust layers at $D\sim700$\,pc are shared between epochs~1 and 2, which means that not only is the short timescale fading of the rings observed, but their evolution can be tracked across several days. Later follow-up observations where taken at 21, 23, and 32 days post-GRB. These observations permit constraints on the evolution of the inner rings of the earlier epochs ($D\sim3000$\,pc in particular). However, these observations are dominated by a variable particle background, so the derived event-list is subject to large systematic uncertainties. Our analysis thus focusses on the first two epochs, particularly the rings shared between epochs, and was performed as follows:

\begin{enumerate}
  \item We reduced the EPIC MOS and pn data in a standard way using SAS~21.0.0 with the most recent calibration files. We employed no time filtering for epoch~1, but excluded the latter half of the second epoch, which had a strong and variable particle background \citep[see][]{Tiengo2023}.
  \item We extracted events in the energy range 0.5--4\,keV in a series of energy bins with bin edges at 0.5, 0.7, 1.0, 1.25, 1.5, 1.75, 2.0, 2.5, 3.0, and 4.0\,keV. The signal peaks around 1.2\,keV with less constraining power towards lower or higher energies.
  \item We extracted effective area and exposure maps and applied these maps to the events data. We removed point sources unrelated to the GRB light and assumed that diffuse light was intrinsically uniform on the sky, which should just be a uniform, constant background in the fit. In Fig.~\ref{fig:EPN-MOS}, we show some exposure-corrected images in different time intervals alongside a difference image to highlight how the rings fade and expand outwards. 
  \item As a function of time post-GRB, $t_d$, and scattering angle, $\theta$, we can make a $(\theta,t_d)$-image (or equivalently a $(D,t_d)$-image or $(D,\theta)$-image given $D=2ct_d\theta^{-2}$, see below). In Fig.~\ref{fig:EPN-MOS} (lower panel), we show how the rings depend on distance (i.e.\ the position of the dust layer, its density and thickness) and angle. In Fig.~\ref{fig:MRN_model} central panel, we show how the data look in $(\theta,t_d)$-space, which is the landscape we use for fitting. 
\end{enumerate}
The width of the point spread functions of the EPIC detectors (with full widths at half maxima, FWHMs, $\lesssim8''$) depend on energy, off-axis angle, and detector-type (i.e.\ MOS or pn). The dominant variation is between detector types,  with the pn having a several tens of percent broader FWHM than the MOS detectors. We account for these PSF functions in our fits. However, the effect of the PSF is negligible in our analysis.
The FWHM increases with increasing off-axis angle in the relevant range from $\theta=0.12^{\mathrm{o}}$ to $\theta=0.22^{\mathrm{o}}$ by up to 10\% for the pn detector, while the variation with energy is subtler, reaching variations at the 5\% level up to the highest energies of 4\,keV for the largest offsets \citep{Read2011}. Note that the early epochs only include MOS data. Including these PSF variations or not has little effect on the measurements presented here because the radial distribution of the dust's radial density is a free parameter in our fit and because our binning is coarser than these arcsecond-scale variations (see Sec.~\ref{sec:radialdust}) -- we model the density of dust at fifty radial distance intervals (logarithmically-spaced between 250\,pc and 2000\,pc), which corresponds to an effective resolution of $15''$, much greater than the sub-arcsec FWHM variation. We have tested this robustness by comparing data with model rings convolved in the angular space with either Gaussian or King functions with the angle-, energy- and detector-dependent FWHMs. This produces similar quality fits and best-fit parameters consistent within 1$\sigma$ in grain-type properties, but makes dust-sheets increasingly narrow and prominent to provide a similar effective width of the rings.  


\subsection{X-ray halo modelling}\label{sec:fitting_formula}
It follows from geometrical arguments that photons scattered on a dust layer travel a further distance than the direct photons from an x-ray source and thereby experience a time delay. Specifically, photons scattered at time $t$ on a single dust layer at distance $D$, with an angle $\theta$, to the central x-ray source exhibit a time-delay $t_d$, with respect to the time of the event, $t_0$ \citep[e.g.][]{Overbeck1965,Xu1986}. 
\begin{equation}
    t_d = t-t_0 = \frac{D}{2c} \theta^2 \label{eq:time_angle_distance}
\end{equation}
Where $c$ is the speed of light and we have assumed the distance to the (Galactic) scattering dust-layer is much smaller than the distance to the extragalactic x-ray source. 

The brightness of the x-ray halo is given by \citep[following the derivation in][]{Beardmore2016}: 
\begin{align}
    S_{\mathrm{halo}}(D, t) = \nonumber \\
    \int_{E_1}^{E_2} \int_{a_{-}}^{a_{+}} \int_{t_1}^{t_2} & A(E) \, F(E, t-t_d) \, n_g(a, D) 
    \frac{D}{t_d} \frac{d\sigma(E, a, \theta)}{d\Omega} \, dt_d \, da \, dE
\end{align}
where $A(E)$ is the effective area of the telescope, $F(E, t)$ is the x-ray source spectrum, which is integrated over the energy and time ranges of interest for the burst, and $d\sigma(E, a, \theta)/d\Omega$ is the differential scattering cross-section. For the dust density along the line-of-sight, $n_g(a, D)$, we assume identical dust grain size and properties for the different Galactic dust scattering layers. Note $n_g(a,D)= dN_g(a,D)/da$ where $N_g(a,D)$ is the density of dust grains with sizes smaller than $a$. Since we assume a single dust grain size distribution for all dust sheets, we can separate $n_g(a,D)$ in distance and grain-size $n_g(a,D)=n(a) n_d(D)$. Here the grain-size distribution is normalised ($\int n(a) da = 1$), while $n_d(D)$ describes the relative abundance of dust at position $D$.
If we use the area-corrected, observed burst fluence, $\mathcal{F}(E) = \int_{t_{1}}^{t_{2}} A(E) F\left(E, t-t_{\mathrm{d}}\right)t^{-1}_{\mathrm{d}}\mathrm{d} t_{\mathrm{d}}$, the integral reduces to: 
\begin{equation}
S_{\mathrm{halo}}\left(E_{\mathrm{b}}, D, t\right)= n_d(D) D \int_{E_{1}}^{E_{2}} \mathcal{F}(E) \int_{a_{-}}^{a^{+}} n(a) \frac{\mathrm{d}\sigma\left(E, a, \theta\right)}{\mathrm{d} \Omega} \mathrm{d} a \mathrm{~d} E
\end{equation}

\subsubsection{The timescale of GRB\,221009A}\label{sec:timescale}
Temporal changes in the spectral properties of the burst GRB\,221009A are not important to our analysis, neither is the X-ray afterglow. First, the short-timescale of the burst \citep[$T_{90}=289\pm1$s, and a much fainter peak several hundred seconds later;][]{Lesage2023} compared to the long timescale of the observations ($t_d$~=~2.31--5.07\,days) means that the fractional timescale is short, i.e.\ $T_{90}/t_d \sim ($7--15$)\times10^{-3}$, or an angular separation of the beginning and end of the burst of $\theta(t_d)-\theta(t_d+T_{90})\lesssim 0.5''$ (given Eq.\,\ref{eq:time_angle_distance}). This is at least an order of magnitude smaller than the resolution of the PSF and the bins over which the model is evaluated. Second, the high energy flux from the GRB itself is two orders of magnitude higher during the prompt phase than the afterglow, suggesting that the contribution of the long-lived afterglow to the X-ray scattered dust emission is small. For the purpose of our analysis, we therefore model GRB\,221009A as a single impulse in time.

Continuous x-ray sources by contrast, necessitate careful modelling of PSF effects to deblend various scattering distances and in addition, longer-lived variable sources require the temporal evolution of the source to be considered as well. On the other hand, short-lived transients like GRB\,221009A scattering on dust-sheets provide a remarkably simple diagnostic system. As the rings fade during their expansion, one merely needs to benchmark the relative fading of lower- and higher-energy photons to directly constrain the properties of the scattering layer.

\subsubsection{Differential scattering cross-section}

We have used anomalous diffraction theory (ADT) to model the scattering. This is applicable across a wide range of grain-sizes and x-ray energies, as it requires only $|m-1|\ll1$ (i.e.\ refraction of photons at entry and exit is negligible), and $a/\lambda\gg1$ \citep[i.e.\ the eikonal approximation,][]{Hulst1957}. Previous work has often used the Rayleigh-Gans (RG) approximation: $(2\pi a/\lambda) |m-1|\ll1$ and absorption is negligible ($k=0$). However, while the bias of using RG \citep{Smith1998} and the improvements from using ADT \citep{Hoffman2016} have been made clear, the downstream implications for x-ray grain-size constraints due to the limitations of using the RG approximation have not previously been examined. In this analysis we employ ADT to obtain robust constraints even for larger grain-sizes and to examine the the optical constants of the grains, $m = n + ik$. 

The computational downside of ADT compared to RG is the lack of a simple analytical form, which instead numerically requires evaluating an integral \citep[which for a spherical grain can be expressed in the following form, see][]{Draine2006}: 

\begin{equation}
\frac{\mathrm{d} \sigma}{\mathrm{d} \Omega} = (2\pi/\lambda)^{2} a^4 \left(\int_0^{1} \left(1-\exp^{i 2 (2\pi a/\lambda) (m-1) \sqrt{1-u^2}} \right) J_0(u 2\pi a/\lambda) u \, du \right)^2
\end{equation}

\noindent This reduces to the RG calculation in the limit of \((2\pi a/\lambda) |m-1|\ll1\), where the differential cross-section for scattering would be \citep[see][]{Mauche1986}:
\begin{equation}
\frac{\mathrm{d} \sigma}{\mathrm{d} \Omega} \propto a^{6} |m-1|^2 \left(\frac{j_{1}(x)}{x}\right)^{2}\left(1+\cos^2\theta\right)
\end{equation}
Where $x = \frac{4\pi a}{\lambda}\sin\left(\frac{\theta}{2}\right)$ and the angular dependence on grain size and photon energy is given by the term $\left(\frac{j_{1}(x)}{x}\right)^{2} (1+\cos^2\theta)$, where $j_{1}$ is the spherical Bessel function of first order. As noted above, the RG approximation assumes that i) the complex refractive index is close to unity $|m-1|\ll1$ and ii) $(2\pi a/\lambda) |m-1|\ll1$, which is more restrictive in the x-ray context. 
For x-ray wavelengths, dust grains predominantly responsible for the scattering are larger than the wavelength of the radiation, i.e.\ $2\pi a/\lambda \gg1$, but even then the latter inequality is generally satisfied because the refractive index is very close to unity, i.e.\ $|m-1|\ll 1$. However, to investigate grain-size distributions and particularly constrain the abundance of larger grains, i.e.\  $0.2-1$\,\micron grains, or examine the impact of various refractive indexes, this falls outside the range of applicability of the RG approximation. Specifically, RG fails towards lower energies ($E\lesssim1\,\mathrm{keV}$ for $a_{+}=0.2$\,\micron) or at even higher energies if larger grains are present \citep{Smith1998}. 

\subsection{Modelling and fitting procedure}

Throughout the analysis we simultaneously fit the radial dust-density distribution (see Sec.~\ref{sec:radialdust}), the grain-size distribution (see Sec.~\ref{sec:dustsize}) and the refractive index (see Sec.~\ref{sec:refractive}). Here the grain properties determine the fading of the rings, while the dust distribution sets the relative intensities of the rings. As our analysis uses the fractional or relative fading of the rings, it is agnostic to the fluence of the original GRB. However, the input spectral shape can be inferred from the observed scattering-ring spectra for specific grain properties (see Sec.~\ref{sec:spectrum}). 

In general, \(\chi^2\) values, such as reported in Table~\ref{table:1}, should be interpreted in the context of the 5922 data points of $S(E_b,D,t)$ modelled. Given the numerous variables encompassing i) radial dust (50--100 parameters), ii) grain size distribution (3--5 parameters) and iii) refractive index (0--4 parameters), this implies typically that the number of degrees of freedom is $\sim5800$ and that a high dimensional parameter space must be surveyed. This exploration is achieved using the \texttt{emcee} MCMC sampler \citep{Foreman-Mackey2013} on the \texttt{lmfit} \texttt{Minimizer} object for $\chi^2$ minimization \citep{Newville2016}, where we run hundreds of walkers over 10\,000 steps (i.e.\ many auto-correlation timescales) to find an optimal minimum, which amounts to several million parameter evaluations of $S_{model}(E_b,D,t)$. We generally assume flat priors on the parameters (beyond assuming positive values of physical quantities). The equilibrium distribution found by the walkers typically display well-defined minima (see Fig.~\ref{fig:posterior_refractive}). The sole exception (where we do not allow a free fit of all parameters) is the case where we enforce the RG approximation (e.g.\ $k=0$, $n-1 \sim0$) to show that this approximation is statistically disfavoured (see rows 1--2 in Table~\ref{table:1}). 

The best-fit reduced chi-squares are $\chi_{\nu}^2\sim1.19-1.42$, i.e.\ substantially above unity. This suggests there is room for further improvements to the data-reduction and likely to the modelling as well. However, the improvements required are not obvious from the fit residuals, with no strong correlation between $\chi^2(E_b, D, t)$ and any particular energy-band, time, or distance (or equivalently offset angle). We speculate that modelling could be improved in future work by i) using late-time imaging for detailed background subtraction of point sources and diffuse light, ii) implementing a higher spatial resolution of dust-sheets and iii) greater resolution or even simultaneous modelling of the landscape in energy, distance and time (instead of discretizing $S_\mathrm{model}(E_b, D, t)$ in bins). Nevertheless, the change in $\Delta\chi^2$ is order a thousand when comparing our best-fit model with free-to-vary $n$ and $k$ to specific mineral compositions, and when comparing grain size distributions (see Table~\ref{table:fit_parameters}), highlighting the strong constraining power of this dataset.  




\begin{figure}[h]
    \centering
    \includegraphics[width=\linewidth,,viewport=0 22 480 329 ,clip=]{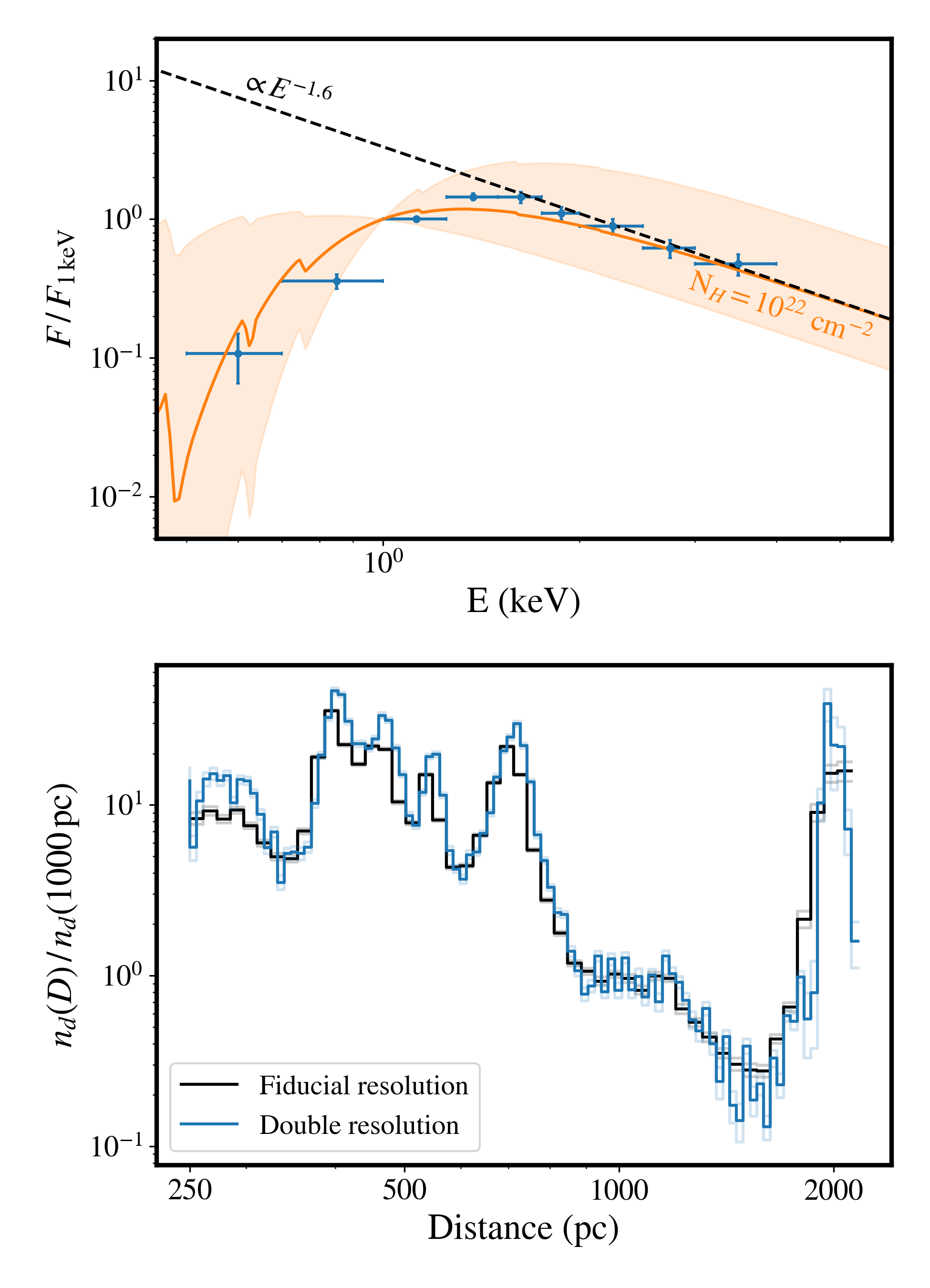}
    \caption{ The dust's radial density landscape fitted for the fiducial computation (black) and for twice as good radial resolution (blue). The increased resolution allows for greater constraints on the radial distribution of dust, but is not important in measuring the energy or azimuthal fading of a ring at a given distance.}
    \label{fig:radial_dust}
\end{figure}

\subsubsection{Radial dust distribution}\label{sec:radialdust}
We parameterise the radial dust distribution $n_d(D)$ at 50 logarithmically-spaced radial bins between 250 and 2200\,pc and evaluated by linear interpolation between these points, which provides similar sheets to those found by \citet{Tiengo2006}. We impose no regularization, yet we find strong correlation between neighbouring points, showing spatial correlations in the line-of-sight extent of the dust sheets. Using cubic-interpolation or doubling the number of evaluated points did not yield significant improvements (see Fig.~\ref{fig:radial_dust}). The dust distribution has little influence on the main thrust of our analysis, which is concerned with how the scattering decreases for a given distance and increasing offset angle. As mentioned previously we also account for energy-, angular- and detector-dependent PSF-effects, which ultimately do not significantly improve the fit or markedly change the best-fit grain-type properties, as the primary effect of including PSF effects is to find somewhat narrower dust sheets in the fit, with little impact on the grain properties. Lastly, we also tried including the smaller offset angle observations from the second epoch (i.e.\ $<0.12^{\mathrm{o}}$), which provide constraints on the dust's radial distribution at greater distances. However, the combination of large distances and the short time baseline probed (relative to the total time post-event) means the angular evolution is smaller, which meant that the grain properties are not more strongly constrained by including all angles. To avoid having even more distance density parameters to fit, we did not include these data in our final analysis.

\subsubsection{Dust grain size distribution}\label{sec:dustsize}
For the dust grain size distribution, we implemented various prescriptions, with our fiducial model being the \citet[MRN]{Mathis1977} distribution, which is characterised by a power-law distribution from the smallest grains, $a_{-}$, to the largest, $a_{+}$. 
\begin{equation}
    n(a) = N_g a^{-q}
\end{equation}

We will also allow for an exponential cut-off for grain-sizes $a > a_+$, with control of the steepness of the cut-off, with $n(a) = N_g a^{-q} f(a,a_c)$ where $f(a,a_c) = \exp(-(a-a_{+})/a_c)$ for $a > a_{+}$ and $f(a)=1$ for $a \leq a_{+}$. For the exponential cut-off version all parameters (e.g.\ $a_c, a_+, q$) are free to vary, but a steep cut-off (small $a_c$) is preferred by the data (see Sec.~\ref{sec:large_grains}). Additionally, to examine the constraining power on large grains we also introduce a lognormal component with free to vary amplitude, centroid and width, i.e.\ not necessarily continuous in the distribution of grain radii with the main grain distribution. 

A more complex distribution is laid out in \citet[]{Weingartner2001}, which has a lognormal distribution for carbon macromolecules and two separate power-law distributions to describe silicate and carbonaceous grains, i.e.\ $n_s(a) = N_{g,s} a^{-q_s}$ with $a_{+,s}$ and $n_c(a) = N_{g,c} a^{-q_c}$ with $a_{+,c}$. In contrast, the recent analysis of \citep{Hensley2023} motivated by polarization data argued for a single composite ``astrodust"-material (for grains larger than $\sim0.02$\,\micron), which, while characterised by a complex size-distribution, notably includes 0.2--0.7\,\micron grains for a mean Milky Way sightline. Therefore, we will also examine the goodness of fit for this grain size distribution model, both with the fiducial grain size distribution from \cite{Hensley2023} and with free to vary parameters. This parameterisation employs various lognormal distributions for small grain sizes (which are however too small to affect scattering rings) alongside a complex fifth-order polynomial to approximate $\log(n(a))$ towards larger grains:
\begin{equation}
    n_{\mathrm{HD}}(a)\propto \exp\left[ \sum_{i=1}^{5} A_i \,\ln\left(\frac{\mathrm{a}}{\angstrom}  \right) \right]
\end{equation}
Where $A_1=-3.40, A_2=-0.807, A_3=0.157, A_4=7.96\times10^{-3}, A_5=-1.68\times10^{-3}$ for the fiducial \citet{Hensley2023} size distribution. We also consider this model, but fit for these polynomial coefficients (labelled HD*, see rows 3--4 in Table\,\ref{table:1}). 

%

\begin{table}[t!]
\caption{Best-fit grain properties and relative incident flux in various bands with 1$\sigma$ uncertainties. The minimum grain-size is provided with a $2\sigma$ upper limit. The flux ratio is reported relative to the flux at 1\,keV and the subscript denotes the lower energy-limit of the band.}
\renewcommand{\arraystretch}{1.4}
\centering
\begin{tabular}{lclc}
\hline \hline
\multicolumn{4}{c}{ {Grain properties and flux-ratio}} \\ \hline
$a_{\mathrm{max}}$        & $0.24 \pm 0.01$     & $\mathrm{F_{0.5\,keV}}$ & $0.09 \pm 0.02$ \\
$a_{\mathrm{min}}$         & <$0.022$            & $\mathrm{F_{0.7\,keV}}$ & $0.34 \pm 0.02$ \\
$q$                                  & $3.08 \pm 0.04$     & $\mathrm{F_{1.25\,keV}}$ & $1.48 \pm 0.04$ \\
&   & $\mathrm{F_{1.5\,keV}}$  & $1.48 \pm 0.06$ \\
$1-n_{1\,\mathrm{keV}}$            & $0.0009 \pm 0.0002$ & $\mathrm{F_{1.75\,keV}}$  & $1.14 \pm 0.06$ \\
$k_{1\,\mathrm{keV}}$                & $0.00027 \pm 0.00007$ & $\mathrm{F_{2\,keV}}$    & $0.91 \pm 0.05$ \\
$k_{\alpha}$                         & $3.4 \pm 0.2$       & $\mathrm{F_{2.5\,keV}}$  & $0.63 \pm 0.05$ \\
$n_{\alpha}$                         & $2.2 \pm 0.2$       & $\mathrm{F_{3\,keV}}$    & $0.48 \pm 0.04$ \\
\hline \hline \\
\end{tabular}
\label{table:grain_properties}

\caption{Dust density $n_d(D)/n_d(1\,\mathrm{kpc})$ at various distances $D$. }
\begin{tabular}{cc@{\hskip 1.5em}cc}
\hline \hline
$D$ [pc] & $n_d(D)/n_d(1000\,\mathrm{pc})$ & $D$ [pc] & $n_d(D)/n_d(1000\,\mathrm{pc})$ \\
\hline
250 & $32 \pm 9$ & 261 & $19 \pm 4$ \\
273 & $21 \pm 4$ & 285 & $22 \pm 3$ \\
298 & $20 \pm 3$ & 311 & $15 \pm 3$ \\
325 & $9.3 \pm 2.0$ & 339 & $8.0 \pm 1.5$ \\
354 & $6.4 \pm 1.1$ & 370 & $9.6 \pm 1.6$ \\
387 & $32 \pm 3$ & 404 & $91 \pm 5$ \\
422 & $51 \pm 4$ & 441 & $39 \pm 3$ \\
460 & $37 \pm 3$ & 481 & $55 \pm 4$ \\
502 & $25 \pm 2$ & 525 & $8.1 \pm 1.2$ \\
548 & $37 \pm 2$ & 572 & $17 \pm 2$ \\
598 & $7.4 \pm 0.9$ & 625 & $10 \pm 1$ \\
652 & $9.6 \pm 0.9$ & 682 & $21 \pm 2$ \\
712 & $37 \pm 2$ & 744 & $42 \pm 2$ \\
777 & $7.0 \pm 1.0$ & 811 & $5.3 \pm 0.6$ \\
848 & $4.1 \pm 0.4$ & 885 & $2.6 \pm 0.3$ \\
925 & $2.8 \pm 0.4$ & 966 & $1.4 \pm 0.3$ \\
1009 & $1.8 \pm 0.3$ & 1054 & $2.8 \pm 0.4$ \\
1101 & $1.4 \pm 0.3$ & 1150 & $1.2 \pm 0.3$ \\
1201 & $2.8 \pm 0.5$ & 1255 & $1.0 \pm 0.2$ \\
1311 & $1.2 \pm 0.2$ & 1369 & $1.4 \pm 0.3$ \\
1430 & $0.92 \pm 0.16$ & 1494 & $0.19 \pm 0.06$ \\
1561 & $0.20 \pm 0.07$ & 1630 & $0.56 \pm 0.13$ \\
1703 & $0.41 \pm 0.10$ & 1779 & $1.2 \pm 0.1$ \\
1858 & $3.4 \pm 0.6$ & 1941 & $44 \pm 9$ \\
2027 & $140 \pm 30$ & 2118 & $20 \pm 5$ \\
\hline \hline
\end{tabular}
\label{table:fit_parameters}
\end{table}

\subsubsection{Refractive index modelling}\label{sec:refractive}
Using the ADT framework, we constrain the best-fit real ($n$) and imaginary ($k$) parts of the refractive index. These can be fit independently at each energy bin or with a spectral model, for example by fitting the energy-dependence of the refractive index with a power-law approximation in energy (i.e.\ $k(E) = k_{\mathrm{1\,keV}} E^{-k_{\alpha}}$ and $1-n(E) = (1-n_{\mathrm{1\,keV}}) E^{-n_{\alpha}}$, where $E$ is the photon energy in\,keV). Such power-law prescriptions are expected to be reasonable approximations for grain-compositions like MgFeSiO$_4$, outside of the localised effects of the atomic absorption edges. We have therefore attempted to constrain the refractive index across the energies where the rings are observed in both early and late epochs (i.e.\ 0.5--4\,keV).  

In addition to parametrizing the real and imaginary components of the refractive index as power-laws, we also tested various linear combinations of astrophysically motivated grain types. In Fig.~\ref{fig:refractive_constraints}, we show the refractive index for various compounds including carbon grains, olivines (i.e.,\ $\mathrm{Mg_2SiO_4, \,MgFeSiO_4, \,Fe_2SiO_4}$) and pyroxenes (e.g.,\     $\mathrm{MgSiO_3}$). By employing linear combinations of these compounds, refractive index with the mass fractional abundance of each grain-type being treated as a free parameter, we can explore potential constraints on the mixing of different compositions  (see rows 5--7 in Table\,\ref{table:1}, or Sec.~\ref{sec:dust_composition}).

\section{Results}
Tables~\ref{table:grain_properties} and \ref{table:fit_parameters} present the best-fit grain properties, the relative intensity in the rings, and the radial dust-density distribution for the best-fit model (e.g.\ MRN with ADT). As mentioned previously, the scattering analysis presented in this paper is not directly sensitive to the fluence of the input GRB as we only consider the energy-dependent fading of the rings with increasing offset angle. Consequently, the radial dust distribution is shown normalized to the value at 1\,kpc and the energy dependence is reported relative to 1\,keV. 

\begin{figure}[h]
    \centering
    \includegraphics[width=\linewidth,,viewport=0 332 480 631 ,clip=]{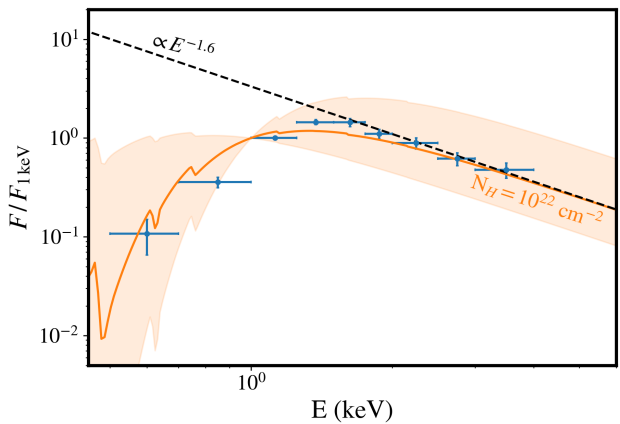}
    \caption{ The input spectrum of the GRB prior to scattering on Galactic dust sheets. The blue errorbars indicate the inferred input x-ray signal, given the best-fit model grain-properties. For comparison the expected shape of a power-law-decline of the prompt signal (black dashed) with substantial MW and host galaxy absorption found in previous work \citep[e.g. $N_\mathrm{H}\sim 3\times 10^{21}$--$2\times 10^{22}$\,cm$^{-2}$;][]{Williams2023,Vaia2025} follow the predicted input spectrum suggested by the best-fit grain-properties. }
    \label{fig:input_spectrum}
\end{figure}

\subsection{GRB prompt spectrum reconstruction}\label{sec:spectrum}
We have allowed complete freedom in the relative intensity of the input flux in the different x-ray energy-bands. As such these energy normalisations do not affect the accuracy of retrieving the other parameters. However, their values are interesting, as they can shed light on the prompt GRB spectrum. The direct light of the GRB itself and the scattered component of the rings both travel through largely the same sheets of dust, resulting in a similar total x-ray absorption. They differ, in that the prompt GRB spectrum is not modified by the energy-dependent efficiency of scattering by the grains. The ring's observed spectrum, given the true grain properties, should thus reproduce the prompt spectrum of the GRB, as it would have been observed in soft x-rays. 

Therefore, given the different constraints on grain properties enumerated below, we can compare how closely our inferred prompt spectrum matches the GRB prompt spectrum. In particular, we see that the input flux in each energy band shows an increase towards higher energies up to 1.5 keV, whereafter the flux decreases as a power-law (Fig.~\ref{fig:input_spectrum}). This flux is reported in units of counts, it is normalised to the bandpass width of the energy bin and we report it relative to the flux at 1\,keV, because we are agnostic to the original fluence of the GRB spectrum. Nevertheless, the inferred input spectrum matches the expectation of a power-law with significant host and Milky Way absorption \cite[e.g.\ $F(E)\propto E^{-1.6}$ with absorption of roughly $N_H\sim10^{22}\,\mathrm{cm^{-2}}$ or see Fig.~5 in][]{Williams2023}. 

Assuming a MW absorption ($N_H\approx7(\pm1)\times10^{21} \mathrm{cm^{-2}}$ from the Galactic dust column on this sight-line) and host absorption ($N_H\approx4\times10^{21} \mathrm{cm^{-2}}$) \citep[e.g.][]{Tiengo2023,Vaia2025}, the derived intrinsic GRB power-law slope is $\Gamma=1.5\pm0.2$ for the best-fit MRN distribution. This fitted incident x-ray spectrum is consistent with the power-law index inferred for the low energy side of the Band model fit to times close to the peak of the prompt emission \citep[e.g.][]{Lesage2023}. Despite this agreement, we still choose to leave the analysis agnostic with complete freedom to vary the spectral shape. For comparison $\Gamma=1.2\pm0.2$ is found in the HD$^*$ model, which might suggest such a model may be disfavoured even more than the relatively larger $\chi^2$ would suggest. However, the spectrum at the peak of the prompt emission in the soft X-ray was not directly measured. It is likely that even stronger constraints on the grain parameters could be obtained in the future by imposing a prior on $F(E)$ based on the observed prompt emission and standard GRB models.

\begin{figure}
    \centering
    \includegraphics[width=\linewidth]{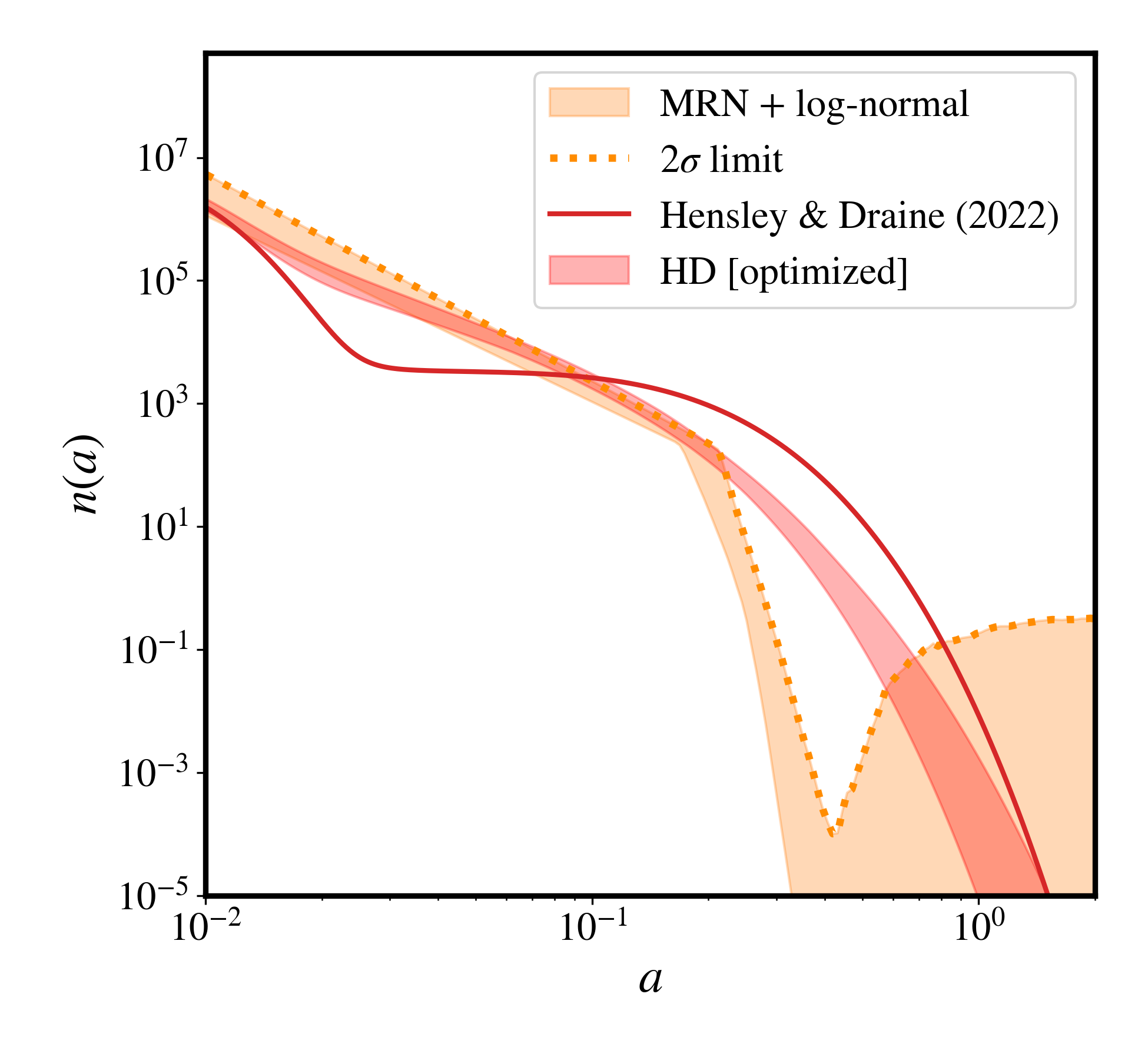}
    \caption{ Dust grain-size distribution, $n(a)$, showing the posterior probability distribution of fits to the GRB\,221009A x-ray rings, with a model based on the MRN \citep{Mathis1977} distribution with an additional large-grain size lognormal component. The proposed grain-size distribution of \citet{Hensley2023} is shown for comparison with the best-fit parameters of this parametrisation shown as well. The dotted line indicates the 95th percentile ($2\sigma$) upper envelope on the grain size distribution. Strong limits can be placed on the population of $a \sim0.4$\,\micron grains, which must be several orders of magnitude less abundant than grains with $a \sim0.2$\,\micron. Because of the lack of grains at this size, the \citet{Hensley2023} grain-size distribution is disfavoured for this sight-line. Larger grains ($\gtrsim1\,$\micron) are not strongly constrained as these predominantly scatter at smaller angles than we measure with these data.}
    \label{fig:grain_size_posterior}
\end{figure}

\subsection{Limits on the population of large grains}\label{sec:large_grains}
Due to the dramatic dependence of the differential scattering cross-section on grain size ($\frac{\mathrm{d} \sigma}{\mathrm{d} \Omega} \propto a^{6}$ for RG, though slightly modified in ADT), any population of large grains is strongly constrained by the angular fading of the rings. 

Fitting a single power-law MRN distribution, we find a maximum grain-size $a_{+}=0.24\pm0.01$. Allowing an exponential cut-off suggests $a_{+}+a_c < 0.28$ at 2$\sigma$ confidence level. 
In Fig.~\ref{fig:grain_size_posterior}, we show the distribution of $n(a)$ inferred from fitting with an MRN grain-size distribution with an additional lognormal distribution (with free to vary amplitude, width and centroid) intended to explore the parameter space of potentially large grains. The resulting 2$\sigma$ upper limit on $n(a)$ is shown, highlighting that while grains larger than $1\,$\micron are not ruled out by the fading rings, grains around $0.4\,$\micron have stringent upper bounds. This strongly suggests that a significant abundance of larger grain sizes is disfavoured along this sightline in the MW. This sightline is thus seemingly in disagreement with the independent lines of evidence, such as modelling in \citet{Hensley2023}, which suggests some population of grains at these sizes. Fitting the landscape of fading rings with a \citet{Hensley2023} distribution (which includes a limited population of 0.3--0.7 micron grains, see Fig.~\ref{fig:grain_size_posterior}) gives a $\chi^2 = 8302$ compared with $\chi^2 = 6950$ for the aforementioned MRN distribution. Allowing the parameters within the parameterisation of \cite{Hensley2023} to be free improves the fit, but not enough to reach the goodness of fit which the more rapid cessation of grains greater than 0.3\,\micron provides. 

The bounds on the grain population around $0.4$\,\micron predominantly comes from the lower energy photons. For instance, 0.5--1\,keV photons forward scatter strongly off 0.2--0.5\,\micron grains at $\sim0.1\degree$, which would fade substantially once the rings expand to $\sim0.2\degree$. However, the rather limited fading of the lower energy photons that we observe rules out families of models with grains of these sizes. The forward scattering from micron-size grains or larger is confined to angles $\theta<0.1\degree$ across the energies analysed here, meaning that we cannot constrain these much larger grains very strongly.

\begin{table}[t]
\caption{ Model comparison using various dust grain-size distributions [\citep[MRN][]{Mathis1977}, \citep[HD][]{Hensley2023}] and various assumptions for refractive index models (e.g.\ ADT or RG).\label{table:table} }
\renewcommand{\arraystretch}{1.4} 
\centering
\begin{tabular}{@{}lcccr@{}}
\hline \hline 
Model & Framework & $a_{+}$ [\textmu m]  & $\chi^2$/DOF & \\ 
\hline \hline 
MRN & ADT  &  $0.235_{-0.006}^{+0.007}$   &  6950/5838  &    \\ %
MRN & RG   &  $0.212_{-0.006}^{+0.008}$   &  8312/5842  &    \\ 
HD  & ADT  &  $\sim0.6$                   &  8302/5841  &   \\  
HD*  & ADT  & $\sim0.5$  &  7925/5836 &  \\  
\hline
MRN & Mg$_{2-x}$Fe$_{x}$SiO$_4$$^\dagger$ &  $0.236 \pm 0.004$   &   7536/5841 &   \\ %
MRN & MgSiO$_3$+Fe   &  $0.241 \pm 0.005$   &  7527/5841  &   \\ %
MRN & MgSiO$_3$+Fe+C   &   $0.236 \pm 0.006$  &  7522/5840  &    \\ %
\hline \hline
\end{tabular}\label{table:1}
\tablefoot{ The HD with fiducial and optimized (HD*) parameters can be seen in Fig.~\ref{fig:grain_size_posterior} (alongside a MRN distribution). The bottom three rows shows similar maximum grain-size and $\chi^2$ for various dust compositions as analysed in Fig.~\ref{fig:Composition}, \ref{fig:MgSiFe_abunds} and \ref{fig:ternary}. $^\dagger$The Fe/Mg ratio is represented by $x/2$, where \(0\leq x\leq2\).}
\end{table}

\begin{figure}
    \centering
    \includegraphics[width=\linewidth]{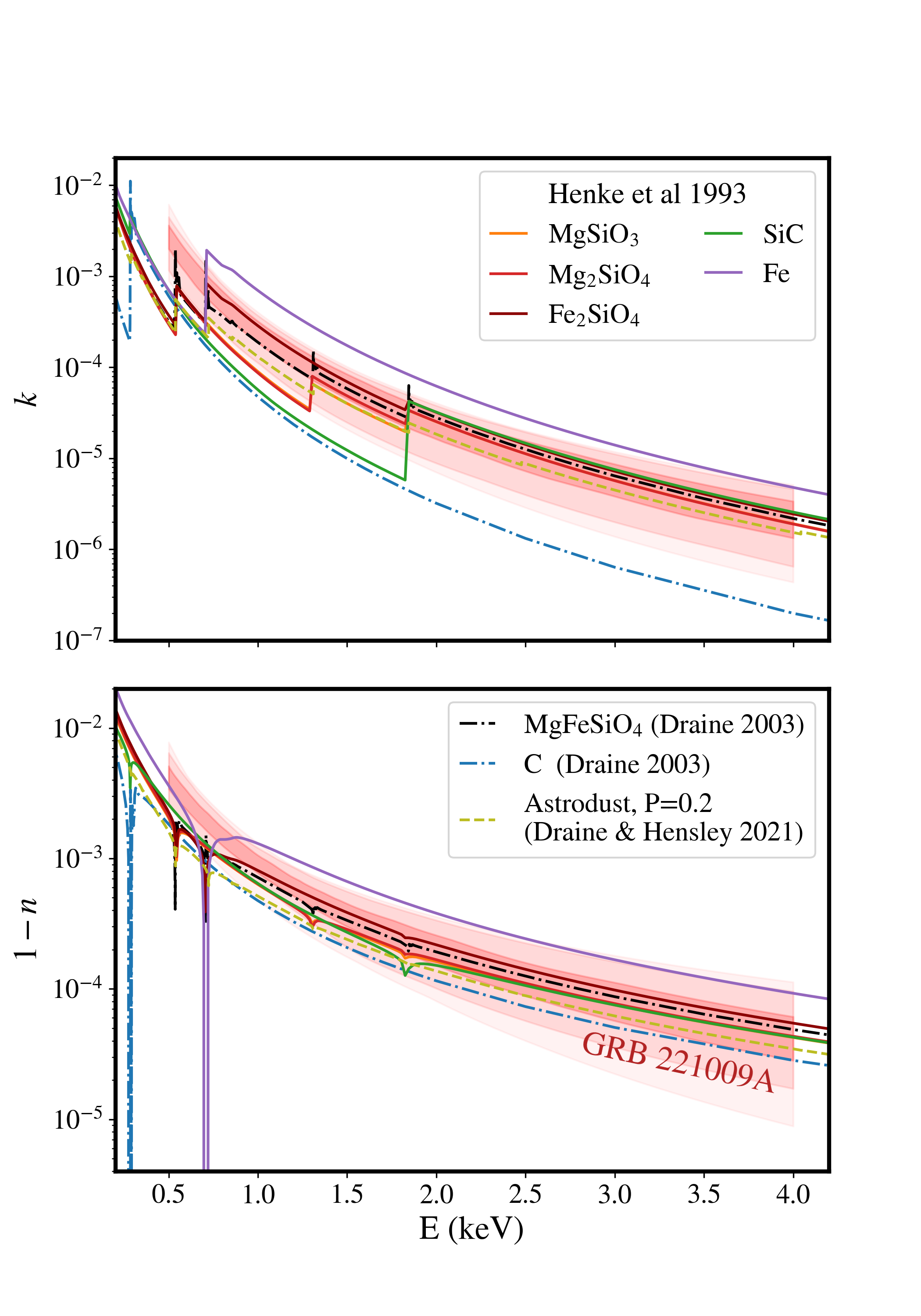}
    \caption{ Imaginary ($k(E)$, top) and real ($n(E)$, bottom) part of the refractive index from fitting photons with energies $0.5\,\mathrm{keV}<E<4\,\mathrm{keV}$. The red line indicates the median expected value from the fading of x-ray rings for GRB\,221009A with shading the indicating 1, 2, and 3$\sigma$ uncertainty regions. The optical constants calculated for different compositions are plotted for comparison \citep{Henke1993,Draine2003} with `astrodust' \citep[with porosity $P=0.2$]{Draine2021}. Individually Fe, SiC, or C don’t match well, while mixtures of Mg- and Fe-silicate grains provide the closest match. }
    \label{fig:refractive_constraints}
\end{figure}

\begin{figure*}
    \centering 
    \includegraphics[width=0.9\linewidth]{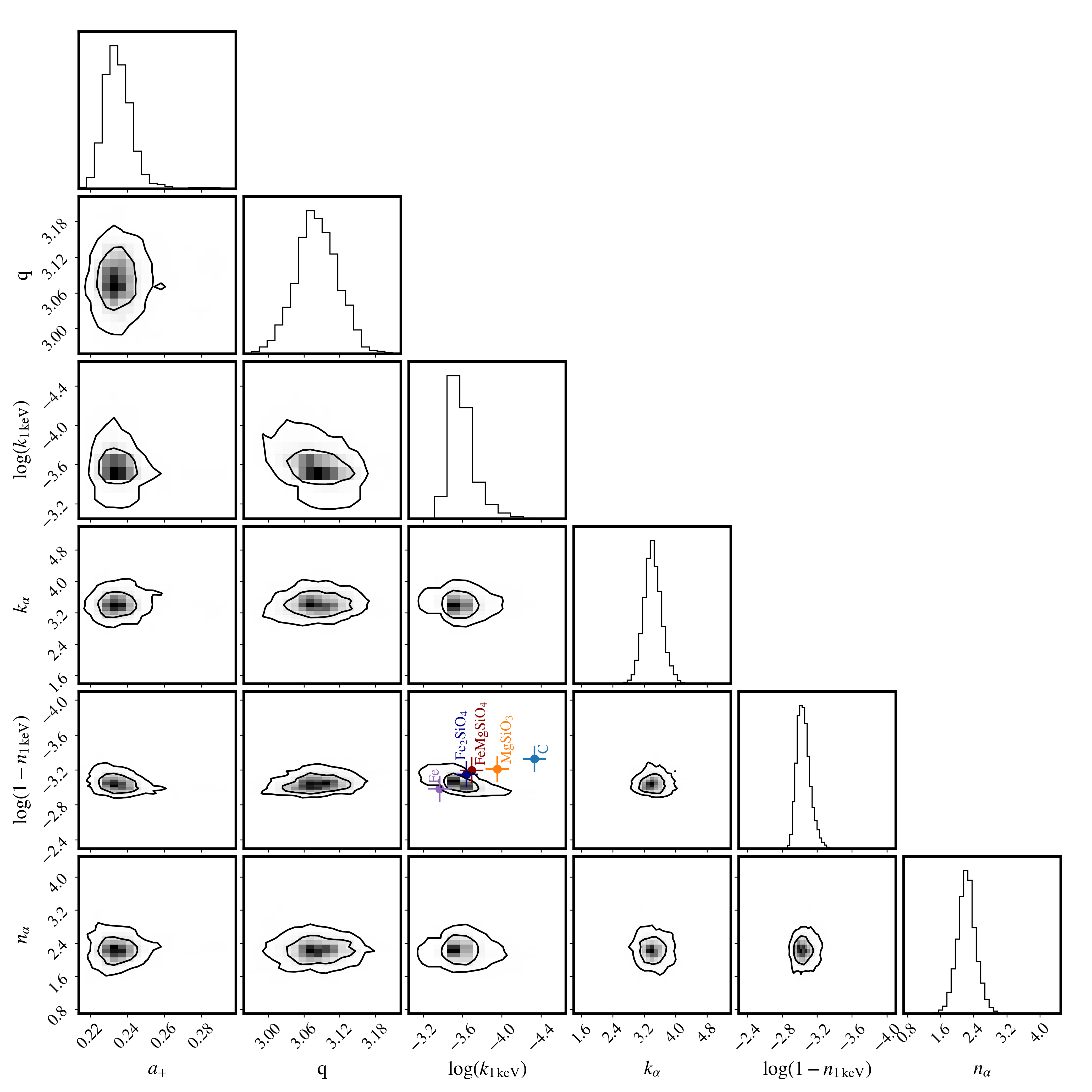} 
    \caption{Corner plot showing the posterior probability distributions of the best fit MRN grain size model with a power-law parametrisation of the refractive index with 1, 2 and 3$\sigma$ contours. The key parameters of the MRN maximum grain-size, $a_{+}$ and power-law slope, $q$, are shown, as are the refractive index's real and imaginary power-law indices, $n_{\alpha}$, $k_{\alpha}$, and values at 1\,keV, $1-n_{\mathrm{1\,keV}}$, $k_{\mathrm{1\,keV}}$. The dots indicate the expected refractive indices for different compositions: carbon grains (blue), metallic Fe grains (purple), Fe$_2$SiO$_4$ (dark blue), MgFeSiO$_4$ (red), and MgSiO$_3$ (orange). While not imposing the Kramers-Kronig relations on the refractive indices, the best-fit $n$ and $k$ are close to standard physical compositions expected for ISM dust. The best-fit parameters are provided in Tables~\ref{table:grain_properties} and \ref{table:fit_parameters}. } 
    \label{fig:posterior_refractive}
\end{figure*}

\subsection{Constraints on the refractive index using anomalous diffraction theory}

In Fig.~\ref{fig:refractive_constraints} we show the resulting constraints on $k(E)$ and $n(E)$ inferred from the fading rings of GRB\,221009A using ADT. The best-fit normalisation in the complex refractive index is  $k_{1\,\mathrm{keV}}= (2.7 \pm 0.7)\times10^{-4}$ and $n_{1\,\mathrm{keV}}= 0.9991 \pm 0.0002$ in linear units. For comparison, the refractive indices from \citet{Henke1993}, \citet{Draine2003}, and \citet{Draine2021} for various dust compounds are shown as a function of energy. For both $k(E)$ and $n(E)$ the normalisation and decline with energy is close to the values expected for MgFeSiO$_4$ or similar Fe-bearing silicates. 
There are some correlations apparent in the data, see Fig.~\ref{fig:posterior_refractive}. Since $k$ sets the absorption within the grain, this makes its effect different from that of the real component, $n$, so that we can constrain them independently -- in the posterior probability landscape $k(E)$ and $n(E)$ are only mildly correlated.

One could tie $k(E)$ and $n(E)$ together by employing the Kramers-Kronig relations, which allows the full complex function to be reconstructed given either the imaginary or real component. In practice this requires an integral over all energies (i.e.\ assuming the properties of $n(E)$/$k(E)$ beyond the x-ray regime). However, there is a reasonable consistency of our inferred ($n(E)$, $k(E)$) from the data with the ($n(E)$, $k(E)$) of various physical compounds, where the latter do include the Kramers-Kronig relations constraint. Our best fits result in optical constants that appear to be close to consistent with the Kramers-Kronig relations without imposing it as a requirement.

\subsection{Dust composition from refractive index constraints}\label{sec:dust_composition}
\begin{figure}
    \centering
    \includegraphics[width=\linewidth]{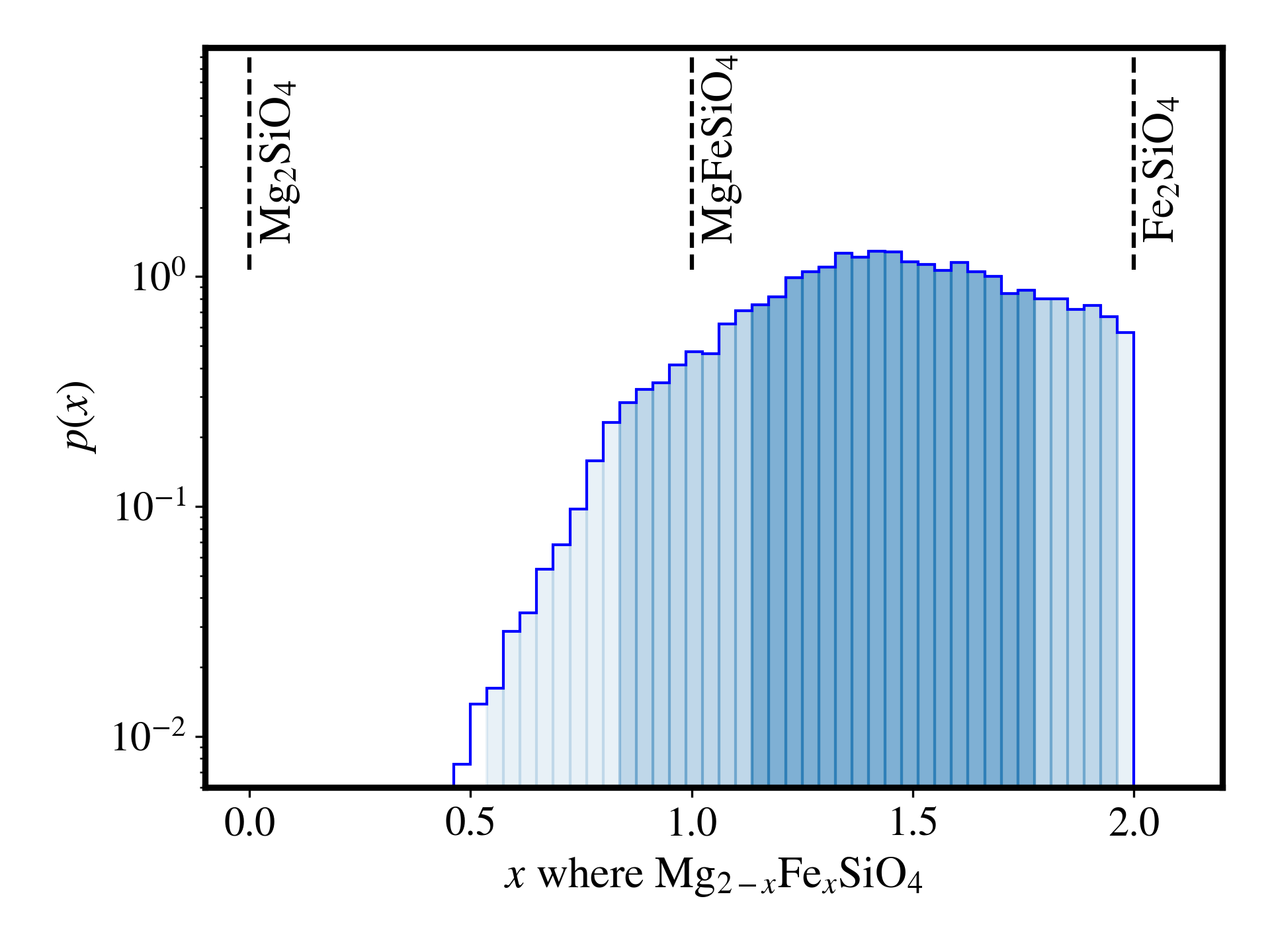}
    \caption{ The posterior probability distribution of the composition of interstellar dust from GRB\,221009A from fitting a limited parameter space with refractive index ranging from \textrm{Mg$_2$SiO$_4$} to \textrm{Fe$_2$SiO$_4$}. The 1, 2, and 3$\sigma$ intervals are indicated with increasingly lighter shades. While \textrm{MgFeSiO$_4$} is consistent with these constraints, a composition dominated by iron (i.e.\ \textrm{Fe$_2$SiO$_4$}) is disfavoured, while an iron-free dust composition is excluded. This seems to be due to the imaginary component of the refractive index, $k$, being too low to be consistent with the data without the presence of a good fraction of some heavier elements, i.e.\ iron, in the dust (Fig.~\ref{fig:refractive_constraints}).
      }
    \label{fig:Composition}
\end{figure}

In light of the strong preference in the data for the refractive index of the iron-rich silicates compared to other compositions, we explore here what constraints can be made compared to a mix of different composition grains, which may be more realistic based on depletion data \citep[e.g.][]{Henning2010}. In Fig.~\ref{fig:Composition}, we show the resulting distribution of dust composition permitted from fitting within the limited parameter space of refractive index between \textrm{Mg$_2$SiO$_4$} and \textrm{Fe$_2$SiO$_4$}. The resulting distribution favours the inclusion of iron in the dust-grains responsible for scattering. MgSiO$_3$ could also be the silicate component in the dust mixture and is perhaps preferred given evidence that the depletion of Mg:Si is near unity and their depletion rates are similar \citep{Jenkins2009,Decia2016,Decia2018}. Thus, 
to compare to the MgFeSiO$_4$, we try 1) a mix of iron-poor silicates with equal Si and Mg (enstatite-composition grains, MgSiO$_3$) and pure Fe grains, and 2) a mix of amorphous carbon, and MgSiO$_3$ and pure Fe grains. We also allow the relative abundances of these components to be a free parameter to see how well they can be constrained. 

\begin{figure}
    \centering
    \includegraphics[width=\linewidth]{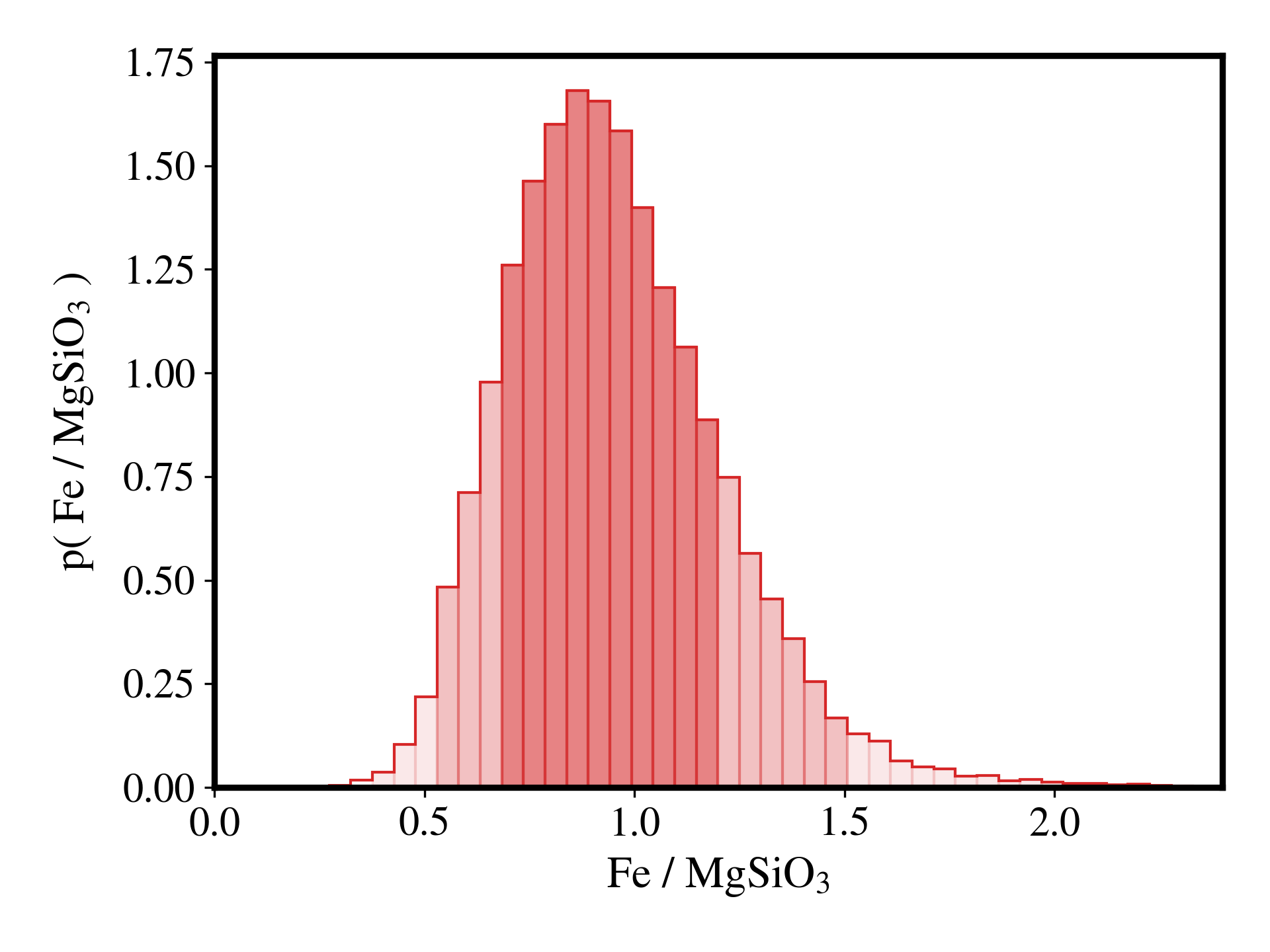}
    \caption{ The posterior probability distribution on the relative composition of Fe and \textrm{MgSiO$_3$}. The 1, 2 and 3$\sigma$ intervals are indicated with increasingly lighter shades. }
    \label{fig:MgSiFe_abunds}
\end{figure}

Overall, we find that we cannot distinguish between MgFeSiO$_4$ and, a mix of MgSiO$_3$ and pure Fe grains, with or without C -- they give similar quality fits (Table~\ref{table:table}). However, beyond this, we find that we require an Fe component. Whether the Fe is chemically part of the silicates or not, we cannot determine here. The Fe should be at a mass quantity comparable to the silicate or carbon components, but the fractional masses of MgSiO$_3$ and carbon are anti-correlated and largely degenerate. 
The relative abundances of model 1) above for MgSiO$_3$ and Fe are shown in Fig.~\ref{fig:MgSiFe_abunds}. The ternary plots and the one-dimensional posterior probabilities for the three component fit (2, above) are shown in Fig.~\ref{fig:ternary}. 

%
\section{Discussion}

\subsection{Bias in the maximum grain-sizes inferred using the Rayleigh-Gans approximation}

The study of x-ray scattering halos has commonly employed the RG approximation to derive constraints on grain size distributions  \citep[e.g.][]{Smith2002, Corrales2015, Beardmore2016, Zhao2024} or dust-column densities \citep[e.g.][]{Pintore2017}. Exploring the larger parameter-landscape of the ADT calculation (which subsumes the RG regime) allows a quantified test of to what extent such constraints are biased by the RG assumptions \citep[see also the discussion in][]{Corrales2015}. 

For instance, \citet{Beardmore2016} assessed the maximum grain size $a_{+}=0.147_{-0.004}^{+0.024}$\,\micron (90\% confidence limits) in the sight-line towards V404 Cygni, while towards Cygnus X-3 \cite{Corrales2015} argued $a_{+}\approx$~0.15--0.2\,\micron. Towards GRB\,221009A, a RG calculation would suggest $a_{+}=0.21\pm0.01$\,\micron from this analysis. However, allowing deviations from RG (i.e.\ freedom in $k$ and $n$) instead gives $a_{+}=0.24\pm0.01$\,\micron. Thus, RG assumptions biased the grain-size downwards, thereby disfavouring the actual best-fit value and highlighting that the systematic uncertainties associated with the RG assumption is larger than the statistical uncertainties of the RG fit. Physically, it is reasonable that the inclusion of absorption (particularly weakening scattering for large grains) would weaken the limit on the large grain population. 

Lastly, it is worth noting that all these analyses are in strong disagreement with the RG fading results of \citep{Zhao2024}, which permit 0.5\,\micron grains from these fading ring of GRB\,221009A. The main difference between these analyses resides in \citet{Zhao2024} integrating over energy, thereby losing the dimension which constrains the grain-size distribution. Specifically, it is the lower energy photons (E<1\,keV) of the rings which do not fade as much and are thus inconsistent with a large grain population.  


\begin{figure}
    \centering
    \includegraphics[width=0.96\linewidth,viewport=1 1 510 530 ,clip=,angle=0]{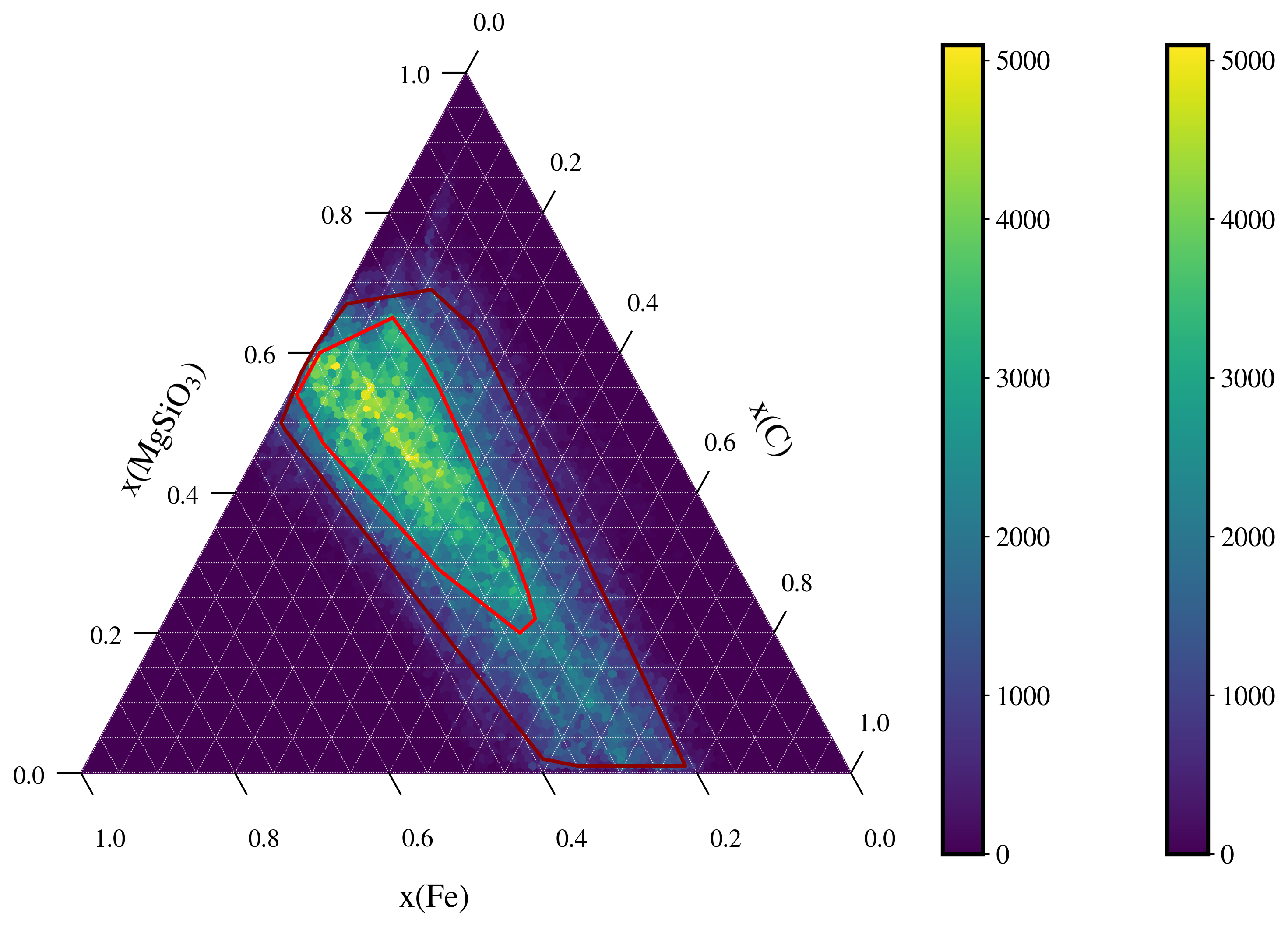}
    \includegraphics[width=\linewidth]{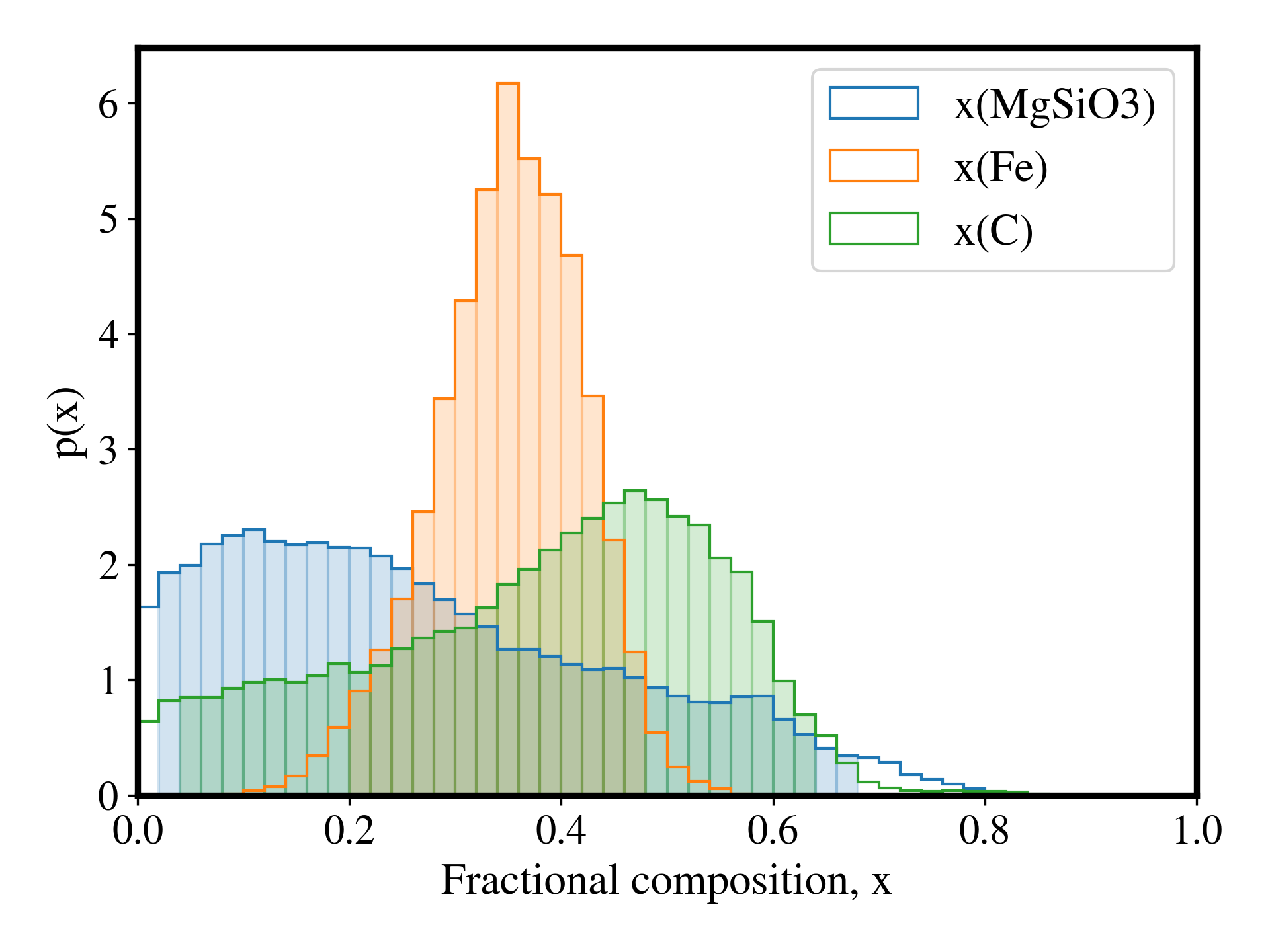} 
    \caption{ Upper panel: Ternary plot with posterior distribution of the mass fractional composition with three representative components: MgSiO$_3$, Fe, and graphitic carbon. Such a three-dimensional space is plotted in a 2D, ternary landscape, because the sum of the contributions equals one. The 1 and 2$\sigma$ contours are indicated with red and dark red lines respectively. Lower panel: The 1D posterior distribution of the relative masses of MgSiO$_3$, Fe, and graphitic carbon. 
    The Fe content is bounded from below and above, $\mathrm{x(Fe)=0.35\pm0.07}$, while an unconstrained ratio mix of C and \textrm{MgSiO$_3$} makes up the rest.
      }
    \label{fig:ternary}
\end{figure}

\subsection{Implications for the chemical composition of dust grains}
The fraction of Fe locked up in dust has been considered in detail \citep[e.g.][]{Poteet2015,Dwek2016,Zeegers2017,Zhukovska2018,Mattsson2019,Westphal2019}. Generally, the Fe depletion out of the gas phase is known to be very high even in the relatively diffuse ISM; in the Galaxy, it is typically depleted at $>90$\%, even in relatively low depletion sightlines \citep{Savage1996,Jenkins2009,Draine2011,Decia2016,Decia2018,Konstantopoulou2022,Konstantopoulou2024}. Exactly where this iron resides in the dust is still an open question. The most likely locations are: chemically bound in silicates, free-floating metallic (or possibly iron oxide) grains, or metallic grains embedded in silicates. Specific environments such as dust-shells surrounding evolved stars display silicates which are generally Mg-rich and Fe-poor \citep{Molster2002} with abundance fraction $\mathrm{A(Fe)\lesssim 1/3\,A(Mg)}$ \citep{Tielens1998}. And in general, most silicates in the ISM are believed to be Mg-rich and iron-poor on the basis both of depletion rate studies \citep{Mattsson2019} and infrared studies of the 9.7\,\micron silicate absorption feature position \citep{Min2007}, though the decisive statement has still to be written.
Fe L-shell extinction data suggest the sightlines to four x-ray binaries are dominated by Mg-rich pyroxene $\mathrm{Mg_{0.75}Fe_{0.25}SiO_3}$ \citep[or olivine,   $\mathrm{MgFeSiO_4}$, for one sight-line][]{Psaradaki2023}. Earlier work on Mg and Si $K$-shell features suggested olivine as the dominant silicate \citep{Zeegers2019,Rogantini2019,Rogantini2020}.
\cite{Corrales2024} found that iron-bearing materials account for 20--30\% of the total mass of interstellar dust from Fe $L$-shell absorption, although notable uncertainties in x-ray atomic databases are present around the Fe $L$-edge \cite{Psaradaki2024}. For reference, we note that the Fe, Si, and Mg abundances by number inferred from the solar photosphere and solar system meteoritics are similar \citep[e.g.\ $\mathrm{A(Fe)/A(Mg) = 0.83 \pm 0.03}$ and $\mathrm{A(Fe)/A(Si) = 0.87 \pm 0.03}$,][]{Asplund2009}, or nearly twice as much in iron by mass, i.e.\ $1.85\pm0.06$ and $1.73\pm 0.06$ respectively, which sums to 0.9 in the mass ratio Fe/(Mg + Si). Accounting for the oxygen believed to be bound in silicates, assuming the Fe is chemically separate, and metallic as a minimum case, this points to a mass ratio of Fe/MgSiO$_3$ of about 0.48. However, this should be considered a lower bound, as gas depletion measurements show that Fe is typically more in the solid form than Si or Mg, which both deplete at similar rates \citep[e.g.][]{Decia2016,Decia2018,Konstantopoulou2022,Konstantopoulou2024}. So that in the diffuse ISM, the expectation is that the mass ratio of Fe/silicates is roughly 0.5--1. Similarly, the C/silicates mass ratio is near unity not accounting for depletion effects, but we caution that the depletion of C is highly uncertain. Assuming a similar depletion for C and Mg or Si, this suggests that we would, a priori, expect the Fe mass fraction in the dust to be approximately 30\%, depending on the depletion level of the ISM.

The total iron mass fraction of the composition for the three species posterior inferred from the scattering rings is substantial (see Fig.\:\ref{fig:ternary}). The median mass fraction, $\mathrm{x(Fe)}=35\pm7$\%, is broadly consistent with the discussion above, particularly at its lower end.
Overall, our compositional constraints should be considered approximate since we do not explore broader ranges of compositions or grain geometries. Improved modelling may be addressed in future work and we return to this in Sect.~\ref{sec:limitations} below.

\subsection{Variation of composition by grain size}\label{sec:grainsize_composition}
Whether the chemical composition of grains varies with grain-size remains an open question. Modelling typically assumes composition is independent of $a$, though there are exceptions \citep[e.g.][]{Jones2017}. It is the largest grains ($\sim$0.1\,\textmu m) that are the dominant contributors to the grain mass\footnote{For an MRN distribution with $q<4$, this follows from $M\propto \int a^{3-q} da$} and also to the scattering signal (as $d\sigma/d\Omega\propto a^{6}$ for RG), so it is particularly these grains whose composition we are most sensitive to. In our analysis we see a consistency between the grain  composition inferred from scattering and that expected from the overall solar elemental composition \citep[i.e.\ comparable magnesium, silicon, and iron by number,][]{Lodders2003}. Thus, these large grains are displaying a composition similar to that expected for the overall elemental composition in most of the Milky Way ISM. 

The spectral resolution of EPIC is quite low, such that individual absorption edges and lines cannot be resolved, as has been done in high spectral resolution studies of direct x-rays absorbed or scattered by the metals in the ISM \citep[e.g.][]{Lee2005}. While the energy and atomic number ($Z$) -dependence of the imaginary component of the refractive index is roughly $k(Z,E)\propto Z^4E^{-3}$\citep{Henke1993}, this energy-dependence is true above the relevant edge energies. Thus, the slope of the resulting $k(E)_{\mathrm{MgFeSiO_4}}$ is less steep than that of any individual element, so we expect a power-law slope $\alpha > -3$. Of the x-ray edges, the prominent oxygen K-edge at 0.532\,keV can abruptly change $k$ by a factor of several. Unfortunately, the signal in the rings of GRB\,230307A drops quickly below 1\,keV \citep{Tiengo2023}, which leaves little statistical constraining power at such low energies. However, the methodology introduced here (and its success at higher energies) does highlight the potential for future studies to constrain the refractive index at lower energies by measuring the scattered light directly at high spectral resolution or better signal-to-noise ratio. Thereby measuring the prominence of the oxygen K-edge and potentially the abundance of oxygen as a function of grain size.


\subsection{Limitations of our scattering framework}\label{sec:limitations}
There are several limitations to our modelling. 
First, the spherical assumption for grains may not hold as explored in \cite{Draine2006}, where oblate spheroids scatter more in the short axis of the grain. For aligned grains, as observed throughout the Milky Way \citep{Andersson2015,PlanckCollaboration2020_XII} this might produce angular dependencies around the ring. Our grain sizes are the radii of spherical grains, so a more detailed modelling would include the possibility of aspherical grains and use the polarisation maps to estimate grain alignment.
Second, substructures within grains -- such as spatial inhomogeneities or porous structures \citep{Hoffman2016} -- may lead to narrowing of the forward scattering peak. 
Third, a single scattering framework is employed, which does not account for the potential multiple scattering if the scattering optical depth becomes large \citep{Draine&Tan2003}. 
Fourth, we do not account for any potential weakening/strengthening of the rings, due to changes in the relative dust column density along sightliness towards larger angles. This may be justified given the random nature of the initial GRB position and the observed minor (azimuthally averaged) radial variation in total hydrogen column density in the sky area around GRB\,221009A \citep{Planck2014,Tiengo2023}. 
Regardless all of these effects may to a greater or lesser extent be partially degenerate with the refractive index in impacting the forward scattering, which perhaps make it all the more remarkable that with a MRN grain-size distribution of spherical grains we find a reasonable refractive index. Finally, we do not allow for the different dust sheets to have different dust properties, and we assume a single grain population with a common composition as a function of size.

While the derived refractive indices are largely in agreement with the Kramers-Kronig relations, the central values of the refractive index parameters do indicate the absorption (i.e.\ $k$) is the one that favours a presence of Fe, whereas the real part (i.e.\ $n$) is more weakly constraining (see Fig.~\ref{fig:refractive_constraints} and Fig.~\ref{fig:posterior_refractive}). Notably, a tension between $n$ and $k$ properties could indicate that the elements mainly responsible for scattering and absorption have different grain size distributions. That is, the freedom of two compositionally distinct grain size distributions may allow a simultaneous modelling of $k$ and $n$, which thus could provide evidence of a varying chemical composition with grain size.



%
\section{Conclusions}

We have analysed the fading x-ray rings, which were observed 2.31--5.07\,days after the brightest GRB ever observed, GRB\,221009A. Our analysis revealed: 
\begin{enumerate}
    \item The lack of strong fading in the rings at lower photon-energies ($E\lesssim1$keV) exclude a significant population of large grains, i.e.\ grain-sizes $a\sim0.4$\,\micron. This rules out a large-grain population along this line of sight, such as those in \citep{Hensley2023}, which are otherwise favoured from analyses designed to reproduce the mean wavelength dependence and polarization of Galactic extinction. 
    \item Implementing anomalous diffraction theory provided statistically superior fits to the often used Rayleigh-Gans approximation, underlining that the latter method is biased due to its assumptions. 
    \item Using anomalous diffraction theory, we measure the complex refractive index, $m(E)=n(E)+ik(E)$, of the grains. The best-fit $n$ and $k$ values show general agreement with the fiducial expectations from depletion studies, approximately consistent with iron and silicates, with $k(E)$ particularly requiring a substantial contribution of iron to the dust grain refractive index. For a mixture of Fe, MgSiO$_3$ and graphitic carbon, we find an iron mass fraction of $0.35_{-0.15}^{+0.12}$ (2$\sigma$ intervals), which, although substantial, is broadly consistent with ISM abundance and depletion data. 
    \item From our fits we also recover the soft x-ray spectrum of the prompt emission that produced the rings. This is consistent with an extrapolation of the low-energy slope of the hard x-ray/$\gamma$-ray emission around the peak time of the prompt emission, even though this is not imposed by the fit.
\end{enumerate}
The analysis presented here has not used all possible available constraints, allowing freedom in most fit parameters. This implies that improved constraints could be attained using more information on the input spectrum, by imposing the Kramers-Kronig relations on the refractive indices, and by using the available data on the dust distribution and polarisation along the line of sight. However, the simplicity of our scattering analysis (e.g.\ we do not look explicitly at porosity, grain geometries, or multiple grain populations) motivates further exploration of this and similar datasets. Such results would further highlight the remarkable insight into dust properties contained within the scattering rings of the brightest extragalactic explosion ever observed, and from bright, transient sources more generally.

\begin{acknowledgements}
The authors would like to thank Bruce Draine for insightful ideas and several important suggestions, including the use of ADT and some questions to pursue. We also thank Anja C. Andersen and the anonymous referee for providing useful feedback on the content. The Cosmic Dawn Center (DAWN) is funded by the Danish National Research Foundation under grant No.~140. AS and DW are funded in part by the European Union (ERC, HEAVYMETAL, 101071865). Views and opinions expressed are, however, those of the authors only and do not necessarily reflect those of the European Union or the European Research Council. Neither the European Union nor the granting authority can be held responsible for them. 
\end{acknowledgements}
\bibliography{refs}

\begin{thebibliography}{68}
\expandafter\ifx\csname natexlab\endcsname\relax\def\natexlab#1{#1}\fi

\bibitem[{{Andersson} {et~al.}(2015){Andersson}, {Lazarian}, \& {Vaillancourt}}]{Andersson2015}
{Andersson}, B.~G., {Lazarian}, A., \& {Vaillancourt}, J.~E. 2015, \araa, 53, 501

\bibitem[{{Asplund} {et~al.}(2009){Asplund}, {Grevesse}, {Sauval}, \& {Scott}}]{Asplund2009}
{Asplund}, M., {Grevesse}, N., {Sauval}, A.~J., \& {Scott}, P. 2009, \araa, 47, 481

\bibitem[{{Beardmore} {et~al.}(2016){Beardmore}, {Willingale}, {Kuulkers}, {Altamirano}, {Motta}, {Osborne}, {Page}, \& {Sivakoff}}]{Beardmore2016}
{Beardmore}, A.~P., {Willingale}, R., {Kuulkers}, E., {et~al.} 2016, \mnras, 462, 1847

\bibitem[{{Corrales} {et~al.}(2024){Corrales}, {Gotthelf}, {Gatuzz}, {Kallman}, {Lee}, {Martins}, {Paerels}, {Psaradaki}, {Schippers}, \& {Savin}}]{Corrales2024}
{Corrales}, L., {Gotthelf}, E.~V., {Gatuzz}, E., {et~al.} 2024, \apj, 965, 172

\bibitem[{{Corrales} \& {Paerels}(2015)}]{Corrales2015}
{Corrales}, L.~R. \& {Paerels}, F. 2015, \mnras, 453, 1121

\bibitem[{{Costantini} \& {Corrales}(2022)}]{Costantini2022}
{Costantini}, E. \& {Corrales}, L. 2022, in Handbook of X-ray and Gamma-ray Astrophysics, ed. C.~{Bambi} \& A.~{Sangangelo}, 40

\bibitem[{{De Cia} {et~al.}(2016){De Cia}, {Ledoux}, {Mattsson}, {Petitjean}, {Srianand}, {Gavignaud}, \& {Jenkins}}]{Decia2016}
{De Cia}, A., {Ledoux}, C., {Mattsson}, L., {et~al.} 2016, \aap, 596, A97

\bibitem[{{De Cia} {et~al.}(2018){De Cia}, {Ledoux}, {Petitjean}, \& {Savaglio}}]{Decia2018}
{De Cia}, A., {Ledoux}, C., {Petitjean}, P., \& {Savaglio}, S. 2018, \aap, 611, A76

\bibitem[{{de Vries} \& {Costantini}(2009)}]{devries2009}
{de Vries}, C.~P. \& {Costantini}, E. 2009, \aap, 497, 393

\bibitem[{{Draine}(2003)}]{Draine2003}
{Draine}, B.~T. 2003, \apj, 598, 1026

\bibitem[{{Draine}(2011)}]{Draine2011}
{Draine}, B.~T. 2011, {Physics of the Interstellar and Intergalactic Medium}

\bibitem[{{Draine} \& {Allaf-Akbari}(2006)}]{Draine2006}
{Draine}, B.~T. \& {Allaf-Akbari}, K. 2006, \apj, 652, 1318

\bibitem[{{Draine} \& {Hensley}(2021)}]{Draine2021}
{Draine}, B.~T. \& {Hensley}, B.~S. 2021, \apj, 909, 94

\bibitem[{{Draine} \& {Tan}(2003)}]{Draine&Tan2003}
{Draine}, B.~T. \& {Tan}, J.~C. 2003, \apj, 594, 347

\bibitem[{{Dwek}(2016)}]{Dwek2016}
{Dwek}, E. 2016, \apj, 825, 136

\bibitem[{{Foreman-Mackey} {et~al.}(2013){Foreman-Mackey}, {Hogg}, {Lang}, \& {Goodman}}]{Foreman-Mackey2013}
{Foreman-Mackey}, D., {Hogg}, D.~W., {Lang}, D., \& {Goodman}, J. 2013, \pasp, 125, 306

\bibitem[{{Heinz} {et~al.}(2015){Heinz}, {Burton}, {Braiding}, {Brandt}, {Jonker}, {Sell}, {Fender}, {Nowak}, \& {Schulz}}]{Heinz2015}
{Heinz}, S., {Burton}, M., {Braiding}, C., {et~al.} 2015, \apj, 806, 265

\bibitem[{{Henke} {et~al.}(1993){Henke}, {Gullikson}, \& {Davis}}]{Henke1993}
{Henke}, B.~L., {Gullikson}, E.~M., \& {Davis}, J.~C. 1993, Atomic Data and Nuclear Data Tables, 54, 181

\bibitem[{{Henning}(2010)}]{Henning2010}
{Henning}, T. 2010, \araa, 48, 21

\bibitem[{{Hensley} \& {Draine}(2023)}]{Hensley2023}
{Hensley}, B.~S. \& {Draine}, B.~T. 2023, \apj, 948, 55

\bibitem[{{Hoffman} \& {Draine}(2016)}]{Hoffman2016}
{Hoffman}, J. \& {Draine}, B.~T. 2016, \apj, 817, 139

\bibitem[{{Jenkins}(2009)}]{Jenkins2009}
{Jenkins}, E.~B. 2009, \apj, 700, 1299

\bibitem[{{Jones} {et~al.}(2017){Jones}, {K{\"o}hler}, {Ysard}, {Bocchio}, \& {Verstraete}}]{Jones2017}
{Jones}, A.~P., {K{\"o}hler}, M., {Ysard}, N., {Bocchio}, M., \& {Verstraete}, L. 2017, \aap, 602, A46

\bibitem[{{Konstantopoulou} {et~al.}(2022){Konstantopoulou}, {De Cia}, {Krogager}, {Ledoux}, {Noterdaeme}, {Fynbo}, {Heintz}, {Watson}, {Andersen}, {Ramburuth-Hurt}, \& {Jermann}}]{Konstantopoulou2022}
{Konstantopoulou}, C., {De Cia}, A., {Krogager}, J.-K., {et~al.} 2022, \aap, 666, A12

\bibitem[{{Konstantopoulou} {et~al.}(2024){Konstantopoulou}, {De Cia}, {Ledoux}, {Krogager}, {Mattsson}, {Watson}, {Heintz}, {P{\'e}roux}, {Noterdaeme}, {Andersen}, {Fynbo}, {Jermann}, \& {Ramburuth-Hurt}}]{Konstantopoulou2024}
{Konstantopoulou}, C., {De Cia}, A., {Ledoux}, C., {et~al.} 2024, \aap, 681, A64

\bibitem[{{Lee} \& {Ravel}(2005)}]{Lee2005}
{Lee}, J.~C. \& {Ravel}, B. 2005, \apj, 622, 970

\bibitem[{{Lesage} {et~al.}(2023){Lesage}, {Veres}, {Briggs}, {Goldstein}, {Kocevski}, {Burns}, {Wilson-Hodge}, {Bhat}, {Huppenkothen}, {Fryer}, {Hamburg}, {Racusin}, {Bissaldi}, {Cleveland}, {Dalessi}, {Fletcher}, {Giles}, {Hristov}, {Hui}, {Mailyan}, {Malacaria}, {Poolakkil}, {Roberts}, {von Kienlin}, {Wood}, {Ajello}, {Arimoto}, {Baldini}, {Ballet}, {Baring}, {Bastieri}, {Gonzalez}, {Bellazzini}, {Bissaldi}, {Blandford}, {Bonino}, {Bruel}, {Buson}, {Cameron}, {Caputo}, {Caraveo}, {Cavazzuti}, {Chiaro}, {Cibrario}, {Ciprini}, {Orestano}, {Crnogorcevic}, {Cuoco}, {Cutini}, {D'Ammando}, {De Gaetano}, {Di Lalla}, {Di Venere}, {Dom{\'\i}nguez}, {Fegan}, {Ferrara}, {Fleischhack}, {Fukazawa}, {Funk}, {Fusco}, {Galanti}, {Gammaldi}, {Gargano}, {Gasbarra}, {Gasparrini}, {Germani}, {Giacchino}, {Giglietto}, {Gill}, {Giroletti}, {Granot}, {Green}, {Grenier}, {Guiriec}, {Gustafsson}, {Hays}, {Hewitt}, {Horan}, {Hou}, {Kuss}, {Latronico}, {Laviron}, {Lemoine-Goumard}, {Li}, {Liodakis}, {Longo}, {Loparco}, {Lorusso},
  {Lovellette}, {Lubrano}, {Maldera}, {Manfreda}, {Mart{\'\i}-Devesa}, {Mazziotta}, {McEnery}, {Mereu}, {Meyer}, {Michelson}, {Mizuno}, {Monzani}, {Morselli}, {Moskalenko}, {Negro}, {Nuss}, {Omodei}, {Orlando}, {Ormes}, {Paneque}, {Panzarini}, {Persic}, {Pesce-Rollins}, {Pillera}, {Piron}, {Poon}, {Porter}, {Principe}, {Rain{\`o}}, {Rando}, {Rani}, {Razzano}, {Razzaque}, {Reimer}, {Reimer}, {Ryde}, {S{\'a}nchez-Conde}, {Parkinson}, {Scotton}, {Serini}, {Sgr{\`o}}, {Sharma}, {Siskind}, {Spandre}, {Spinelli}, {Tajima}, {Torres}, {Valverde}, {Venters}, {Wadiasingh}, {Wood}, \& {Zaharijas}}]{Lesage2023}
{Lesage}, S., {Veres}, P., {Briggs}, M.~S., {et~al.} 2023, \apjl, 952, L42

\bibitem[{{Lodders}(2003)}]{Lodders2003}
{Lodders}, K. 2003, \apj, 591, 1220

\bibitem[{{Malesani} {et~al.}(2023){Malesani}, {Levan}, {Izzo}, {de Ugarte Postigo}, {Ghirlanda}, {Heintz}, {Kann}, {Lamb}, {Palmerio}, {Salafia}, {Salvaterra}, {Tanvir}, {Ag{\"u}{\'\i} Fern{\'a}ndez}, {Campana}, {Chrimes}, {D'Avanzo}, {D'Elia}, {Della Valle}, {De Pasquale}, {Fynbo}, {Gaspari}, {Gompertz}, {Hartmann}, {Hjorth}, {Jakobsson}, {Palazzi}, {Pian}, {Pugliese}, {Ravasio}, {Rossi}, {Saccardi}, {Schady}, {Schneider}, {Sollerman}, {Starling}, {Th{\"o}ne}, {van der Horst}, {Vergani}, {Watson}, {Wiersema}, {Xu}, \& {Zafar}}]{Malesani2023}
{Malesani}, D.~B., {Levan}, A.~J., {Izzo}, L., {et~al.} 2023, arXiv e-prints, arXiv:2302.07891

\bibitem[{{Mathis} {et~al.}(1977){Mathis}, {Rumpl}, \& {Nordsieck}}]{Mathis1977}
{Mathis}, J.~S., {Rumpl}, W., \& {Nordsieck}, K.~H. 1977, \apj, 217, 425

\bibitem[{{Mattsson} {et~al.}(2019){Mattsson}, {De Cia}, {Andersen}, \& {Petitjean}}]{Mattsson2019}
{Mattsson}, L., {De Cia}, A., {Andersen}, A.~C., \& {Petitjean}, P. 2019, \aap, 624, A103

\bibitem[{{Mauche} \& {Gorenstein}(1986)}]{Mauche1986}
{Mauche}, C.~W. \& {Gorenstein}, P. 1986, \apj, 302, 371

\bibitem[{{Min} {et~al.}(2007){Min}, {Waters}, {de Koter}, {Hovenier}, {Keller}, \& {Markwick-Kemper}}]{Min2007}
{Min}, M., {Waters}, L.~B.~F.~M., {de Koter}, A., {et~al.} 2007, \aap, 462, 667

\bibitem[{{Molster} {et~al.}(2002){Molster}, {Waters}, \& {Tielens}}]{Molster2002}
{Molster}, F.~J., {Waters}, L.~B.~F.~M., \& {Tielens}, A.~G.~G.~M. 2002, \aap, 382, 222

\bibitem[{{Newville} {et~al.}(2016){Newville}, {Stensitzki}, {Allen}, {Rawlik}, {Ingargiola}, \& {Nelson}}]{Newville2016}
{Newville}, M., {Stensitzki}, T., {Allen}, D.~B., {et~al.} 2016, {Lmfit: Non-Linear Least-Square Minimization and Curve-Fitting for Python}, Astrophysics Source Code Library, record ascl:1606.014

\bibitem[{{Overbeck}(1965)}]{Overbeck1965}
{Overbeck}, J.~W. 1965, \apj, 141, 864

\bibitem[{{Pintore} {et~al.}(2017){Pintore}, {Tiengo}, {Mereghetti}, {Vianello}, {Salvaterra}, {Esposito}, {Costantini}, {Giuliani}, \& {Bosnjak}}]{Pintore2017}
{Pintore}, F., {Tiengo}, A., {Mereghetti}, S., {et~al.} 2017, \mnras, 472, 1465

\bibitem[{{Planck Collaboration} {et~al.}(2014){Planck Collaboration}, {Abergel}, {Ade}, {Aghanim}, {Alves}, {Aniano}, {Armitage-Caplan}, {Arnaud}, {Ashdown}, {Atrio-Barandela}, {Aumont}, {Baccigalupi}, {Banday}, {Barreiro}, {Bartlett}, {Battaner}, {Benabed}, {Beno{\^\i}t}, {Benoit-L{\'e}vy}, {Bernard}, {Bersanelli}, {Bielewicz}, {Bobin}, {Bock}, {Bonaldi}, {Bond}, {Borrill}, {Bouchet}, {Boulanger}, {Bridges}, {Bucher}, {Burigana}, {Butler}, {Cardoso}, {Catalano}, {Chamballu}, {Chary}, {Chiang}, {Chiang}, {Christensen}, {Church}, {Clemens}, {Clements}, {Colombi}, {Colombo}, {Combet}, {Couchot}, {Coulais}, {Crill}, {Curto}, {Cuttaia}, {Danese}, {Davies}, {Davis}, {de Bernardis}, {de Rosa}, {de Zotti}, {Delabrouille}, {Delouis}, {D{\'e}sert}, {Dickinson}, {Diego}, {Dole}, {Donzelli}, {Dor{\'e}}, {Douspis}, {Draine}, {Dupac}, {Efstathiou}, {En{\ss}lin}, {Eriksen}, {Falgarone}, {Finelli}, {Forni}, {Frailis}, {Fraisse}, {Franceschi}, {Galeotta}, {Ganga}, {Ghosh}, {Giard}, {Giardino}, {Giraud-H{\'e}raud},
  {Gonz{\'a}lez-Nuevo}, {G{\'o}rski}, {Gratton}, {Gregorio}, {Grenier}, {Gruppuso}, {Guillet}, {Hansen}, {Hanson}, {Harrison}, {Helou}, {Henrot-Versill{\'e}}, {Hern{\'a}ndez-Monteagudo}, {Herranz}, {Hildebrandt}, {Hivon}, {Hobson}, {Holmes}, {Hornstrup}, {Hovest}, {Huffenberger}, {Jaffe}, {Jaffe}, {Jewell}, {Joncas}, {Jones}, {Juvela}, {Keih{\"a}nen}, {Keskitalo}, {Kisner}, {Knoche}, {Knox}, {Kunz}, {Kurki-Suonio}, {Lagache}, {L{\"a}hteenm{\"a}ki}, {Lamarre}, {Lasenby}, {Laureijs}, {Lawrence}, {Leonardi}, {Le{\'o}n-Tavares}, {Lesgourgues}, {Levrier}, {Liguori}, {Lilje}, {Linden-V{\o}rnle}, {L{\'o}pez-Caniego}, {Lubin}, {Mac{\'\i}as-P{\'e}rez}, {Maffei}, {Maino}, {Mandolesi}, {Maris}, {Marshall}, {Martin}, {Mart{\'\i}nez-Gonz{\'a}lez}, {Masi}, {Massardi}, {Matarrese}, {Matthai}, {Mazzotta}, {McGehee}, {Melchiorri}, {Mendes}, {Mennella}, {Migliaccio}, {Mitra}, {Miville-Desch{\^e}nes}, {Moneti}, {Montier}, {Morgante}, {Mortlock}, {Munshi}, {Murphy}, {Naselsky}, {Nati}, {Natoli}, {Netterfield},
  {N{\o}rgaard-Nielsen}, {Noviello}, {Novikov}, {Novikov}, {Osborne}, {Oxborrow}, {Paci}, {Pagano}, {Pajot}, {Paladini}, {Paoletti}, {Pasian}, {Patanchon}, {Perdereau}, {Perotto}, {Perrotta}, {Piacentini}, {Piat}, {Pierpaoli}, {Pietrobon}, {Plaszczynski}, {Pointecouteau}, {Polenta}, {Ponthieu}, {Popa}, {Poutanen}, {Pratt}, {Pr{\'e}zeau}, {Prunet}, {Puget}, {Rachen}, {Reach}, {Rebolo}, {Reinecke}, {Remazeilles}, {Renault}, {Ricciardi}, \& {Riller}}]{Planck2014}
{Planck Collaboration}, {Abergel}, A., {Ade}, P.~A.~R., {et~al.} 2014, \aap, 571, A11

\bibitem[{{Planck Collaboration} {et~al.}(2020){Planck Collaboration}, {Aghanim}, {Akrami}, {Alves}, {Ashdown}, {Aumont}, {Baccigalupi}, {Ballardini}, {Banday}, {Barreiro}, {Bartolo}, {Basak}, {Benabed}, {Bernard}, {Bersanelli}, {Bielewicz}, {Bock}, {Bond}, {Borrill}, {Bouchet}, {Boulanger}, {Bracco}, {Bucher}, {Burigana}, {Calabrese}, {Cardoso}, {Carron}, {Chary}, {Chiang}, {Colombo}, {Combet}, {Crill}, {Cuttaia}, {de Bernardis}, {de Zotti}, {Delabrouille}, {Delouis}, {Di Valentino}, {Dickinson}, {Diego}, {Dor{\'e}}, {Douspis}, {Ducout}, {Dupac}, {Efstathiou}, {Elsner}, {En{\ss}lin}, {Eriksen}, {Falgarone}, {Fantaye}, {Fernandez-Cobos}, {Ferri{\`e}re}, {Finelli}, {Forastieri}, {Frailis}, {Fraisse}, {Franceschi}, {Frolov}, {Galeotta}, {Galli}, {Ganga}, {G{\'e}nova-Santos}, {Gerbino}, {Ghosh}, {Gonz{\'a}lez-Nuevo}, {G{\'o}rski}, {Gratton}, {Green}, {Gruppuso}, {Gudmundsson}, {Guillet}, {Handley}, {Hansen}, {Helou}, {Herranz}, {Hivon}, {Huang}, {Jaffe}, {Jones}, {Keih{\"a}nen}, {Keskitalo}, {Kiiveri}, {Kim},
  {Krachmalnicoff}, {Kunz}, {Kurki-Suonio}, {Lagache}, {Lamarre}, {Lasenby}, {Lattanzi}, {Lawrence}, {Le Jeune}, {Levrier}, {Liguori}, {Lilje}, {Lindholm}, {L{\'o}pez-Caniego}, {Lubin}, {Ma}, {Mac{\'\i}as-P{\'e}rez}, {Maggio}, {Maino}, {Mandolesi}, {Mangilli}, {Marcos-Caballero}, {Maris}, {Martin}, {Mart{\'\i}nez-Gonz{\'a}lez}, {Matarrese}, {Mauri}, {McEwen}, {Melchiorri}, {Mennella}, {Migliaccio}, {Miville-Desch{\^e}nes}, {Molinari}, {Moneti}, {Montier}, {Morgante}, {Moss}, {Natoli}, {Pagano}, {Paoletti}, {Patanchon}, {Perrotta}, {Pettorino}, {Piacentini}, {Polastri}, {Polenta}, {Puget}, {Rachen}, {Reinecke}, {Remazeilles}, {Renzi}, {Ristorcelli}, {Rocha}, {Rosset}, {Roudier}, {Rubi{\~n}o-Mart{\'\i}n}, {Ruiz-Granados}, {Salvati}, {Sandri}, {Savelainen}, {Scott}, {Sirignano}, {Sunyaev}, {Suur-Uski}, {Tauber}, {Tavagnacco}, {Tenti}, {Toffolatti}, {Tomasi}, {Trombetti}, {Valiviita}, {Vansyngel}, {Van Tent}, {Vielva}, {Villa}, {Vittorio}, {Wandelt}, {Wehus}, {Zacchei}, \& {Zonca}}]{PlanckCollaboration2020_XII}
{Planck Collaboration}, {Aghanim}, N., {Akrami}, Y., {et~al.} 2020, \aap, 641, A12

\bibitem[{{Poteet} {et~al.}(2015){Poteet}, {Whittet}, \& {Draine}}]{Poteet2015}
{Poteet}, C.~A., {Whittet}, D. C.~B., \& {Draine}, B.~T. 2015, \apj, 801, 110

\bibitem[{{Predehl} \& {Schmitt}(1995)}]{Predehl1995}
{Predehl}, P. \& {Schmitt}, J.~H.~M.~M. 1995, \aap, 293, 889

\bibitem[{{Psaradaki} {et~al.}(2024){Psaradaki}, {Corrales}, {Werk}, {Jensen}, {Costantini}, {Mehdipour}, {Cilley}, {Schulz}, {Kaastra}, {Garc{\'\i}a}, {Valencic}, {Kallman}, \& {Paerels}}]{Psaradaki2024}
{Psaradaki}, I., {Corrales}, L., {Werk}, J., {et~al.} 2024, \aj, 167, 217

\bibitem[{{Psaradaki} {et~al.}(2023){Psaradaki}, {Costantini}, {Rogantini}, {Mehdipour}, {Corrales}, {Zeegers}, {de Groot}, {den Herder}, {Mutschke}, {Trasobares}, {de Vries}, \& {Waters}}]{Psaradaki2023}
{Psaradaki}, I., {Costantini}, E., {Rogantini}, D., {et~al.} 2023, \aap, 670, A30

\bibitem[{{Read} {et~al.}(2011){Read}, {Rosen}, {Saxton}, \& {Ramirez}}]{Read2011}
{Read}, A.~M., {Rosen}, S.~R., {Saxton}, R.~D., \& {Ramirez}, J. 2011, \aap, 534, A34

\bibitem[{{Rogantini} {et~al.}(2019){Rogantini}, {Costantini}, {Zeegers}, {de Vries}, {Mehdipour}, {de Groot}, {Mutschke}, {Psaradaki}, \& {Waters}}]{Rogantini2019}
{Rogantini}, D., {Costantini}, E., {Zeegers}, S.~T., {et~al.} 2019, \aap, 630, A143

\bibitem[{{Rogantini} {et~al.}(2020){Rogantini}, {Costantini}, {Zeegers}, {Mehdipour}, {Psaradaki}, {Raassen}, {de Vries}, \& {Waters}}]{Rogantini2020}
{Rogantini}, D., {Costantini}, E., {Zeegers}, S.~T., {et~al.} 2020, \aap, 641, A149

\bibitem[{{Rolf}(1983)}]{Rolf1983}
{Rolf}, D.~P. 1983, \nat, 302, 46

\bibitem[{{Savage} \& {Sembach}(1996)}]{Savage1996}
{Savage}, B.~D. \& {Sembach}, K.~R. 1996, \araa, 34, 279

\bibitem[{{Smith} \& {Dwek}(1998)}]{Smith1998}
{Smith}, R.~K. \& {Dwek}, E. 1998, \apj, 503, 831

\bibitem[{{Smith} {et~al.}(2002){Smith}, {Edgar}, \& {Shafer}}]{Smith2002}
{Smith}, R.~K., {Edgar}, R.~J., \& {Shafer}, R.~A. 2002, \apj, 581, 562

\bibitem[{{Tielens} {et~al.}(1998){Tielens}, {Waters}, {Molster}, \& {Justtanont}}]{Tielens1998}
{Tielens}, A.~G.~G.~M., {Waters}, L.~B.~F.~M., {Molster}, F.~J., \& {Justtanont}, K. 1998, \apss, 255, 415

\bibitem[{{Tiengo} \& {Mereghetti}(2006)}]{Tiengo2006}
{Tiengo}, A. \& {Mereghetti}, S. 2006, \aap, 449, 203

\bibitem[{{Tiengo} {et~al.}(2023){Tiengo}, {Pintore}, {Vaia}, {Filippi}, {Sacchi}, {Esposito}, {Rigoselli}, {Mereghetti}, {Salvaterra}, {{\v{S}}iljeg}, {Bracco}, {Bo{\v{s}}njak}, {Jeli{\'c}}, \& {Campana}}]{Tiengo2023}
{Tiengo}, A., {Pintore}, F., {Vaia}, B., {et~al.} 2023, \apjl, 946, L30

\bibitem[{{Tiengo} {et~al.}(2010){Tiengo}, {Vianello}, {Esposito}, {Mereghetti}, {Giuliani}, {Costantini}, {Israel}, {Stella}, {Turolla}, {Zane}, {Rea}, {G{\"o}tz}, {Bernardini}, {Moretti}, {Romano}, {Ehle}, \& {Gehrels}}]{Tiengo2010}
{Tiengo}, A., {Vianello}, G., {Esposito}, P., {et~al.} 2010, \apj, 710, 227

\bibitem[{{Vaia} {et~al.}(2025){Vaia}, {Bo{\v{s}}njak}, {Bracco}, {Campana}, {Esposito}, {Jeli{\'c}}, {Sacchi}, \& {Tiengo}}]{Vaia2025}
{Vaia}, B., {Bo{\v{s}}njak}, {\v{Z}}., {Bracco}, A., {et~al.} 2025, \aap, 696, A9

\bibitem[{{van de Hulst}(1957)}]{Hulst1957}
{van de Hulst}, H.~C. 1957, {Light Scattering by Small Particles}

\bibitem[{{Vasilopoulos} {et~al.}(2023){Vasilopoulos}, {Karavola}, {Stathopoulos}, \& {Petropoulou}}]{Vasilopoulos2023}
{Vasilopoulos}, G., {Karavola}, D., {Stathopoulos}, S.~I., \& {Petropoulou}, M. 2023, \mnras, 521, 1590

\bibitem[{{Vaughan} {et~al.}(2004){Vaughan}, {Willingale}, {O'Brien}, {Osborne}, {Reeves}, {Levan}, {Watson}, {Tedds}, {Watson}, {Santos-Lle{\'o}}, {Rodr{\'\i}guez-Pascual}, \& {Schartel}}]{Vaughan2004}
{Vaughan}, S., {Willingale}, R., {O'Brien}, P.~T., {et~al.} 2004, \apjl, 603, L5

\bibitem[{{Vaughan} {et~al.}(2006){Vaughan}, {Willingale}, {Romano}, {Osborne}, {Goad}, {Beardmore}, {Burrows}, {Campana}, {Chincarini}, {Covino}, {Moretti}, {O'Brien}, {Page}, {Supper}, \& {Tagliaferri}}]{Vaughan2006}
{Vaughan}, S., {Willingale}, R., {Romano}, P., {et~al.} 2006, \apj, 639, 323

\bibitem[{{Watson} {et~al.}(2006){Watson}, {Vaughan}, {Willingale}, {Hjorth}, {Foley}, {Fynbo}, {Jakobsson}, {Levan}, {O'Brien}, {Osborne}, {Pedersen}, {Reeves}, {Tedds}, \& {Watson}}]{Watson2006}
{Watson}, D., {Vaughan}, S.~A., {Willingale}, R., {et~al.} 2006, \apj, 636, 967

\bibitem[{{Weingartner} \& {Draine}(2001)}]{Weingartner2001}
{Weingartner}, J.~C. \& {Draine}, B.~T. 2001, \apj, 548, 296

\bibitem[{{Westphal} {et~al.}(2019){Westphal}, {Butterworth}, {Tomsick}, \& {Gainsforth}}]{Westphal2019}
{Westphal}, A.~J., {Butterworth}, A.~L., {Tomsick}, J.~A., \& {Gainsforth}, Z. 2019, \apj, 872, 66

\bibitem[{{Williams} {et~al.}(2023){Williams}, {Kennea}, {Dichiara}, {Kobayashi}, {Iwakiri}, {Beardmore}, {Evans}, {Heinz}, {Lien}, {Oates}, {Negoro}, {Cenko}, {Buisson}, {Hartmann}, {Jaisawal}, {Kuin}, {Lesage}, {Page}, {Parsotan}, {Pasham}, {Sbarufatti}, {Siegel}, {Sugita}, {Younes}, {Ambrosi}, {Arzoumanian}, {Bernardini}, {Campana}, {Capalbi}, {Caputo}, {D'A{\`\i}}, {D'Avanzo}, {D'Elia}, {De Pasquale}, {Eyles-Ferris}, {Ferrara}, {Gendreau}, {Gropp}, {Kawai}, {Klingler}, {Laha}, {Melandri}, {Mihara}, {Moss}, {O'Brien}, {Osborne}, {Palmer}, {Perri}, {Serino}, {Sonbas}, {Stamatikos}, {Starling}, {Tagliaferri}, {Tohuvavohu}, {Zane}, \& {Ziaeepour}}]{Williams2023}
{Williams}, M.~A., {Kennea}, J.~A., {Dichiara}, S., {et~al.} 2023, \apjl, 946, L24

\bibitem[{{Xu} {et~al.}(1986){Xu}, {McCray}, \& {Kelley}}]{Xu1986}
{Xu}, Y., {McCray}, R., \& {Kelley}, R. 1986, \nat, 319, 652

\bibitem[{{Zeegers} {et~al.}(2017){Zeegers}, {Costantini}, {de Vries}, {Tielens}, {Chihara}, {de Groot}, {Mutschke}, {Waters}, \& {Zeidler}}]{Zeegers2017}
{Zeegers}, S.~T., {Costantini}, E., {de Vries}, C.~P., {et~al.} 2017, \aap, 599, A117

\bibitem[{{Zeegers} {et~al.}(2019){Zeegers}, {Costantini}, {Rogantini}, {de Vries}, {Mutschke}, {Mohr}, {de Groot}, \& {Tielens}}]{Zeegers2019}
{Zeegers}, S.~T., {Costantini}, E., {Rogantini}, D., {et~al.} 2019, \aap, 627, A16

\bibitem[{{Zhao} \& {Shen}(2024)}]{Zhao2024}
{Zhao}, G. \& {Shen}, R.-F. 2024, \apj, 970, 124

\bibitem[{{Zhukovska} {et~al.}(2018){Zhukovska}, {Henning}, \& {Dobbs}}]{Zhukovska2018}
{Zhukovska}, S., {Henning}, T., \& {Dobbs}, C. 2018, \apj, 857, 94

\end{thebibliography}
\bibliographystyle{aa}

\appendix

\end{document}